\documentclass[aps,nofootinbib,showpacs,floats,letterpaper,floatfix,groupedaddress,eqsecnum]{revtex4}

\usepackage{dcolumn,epsfig}
\usepackage{amssymb,amsmath}

\def\be{\begin{equation}}
\def\ee{\end{equation}}
\def\beq{\begin{eqnarray}}
\def\eeq{\end{eqnarray}}

\def\IL{\relax{\rm I\kern-.18em L}}

\def\f{\frac}

\begin{document}

\title{Matched-filtering and parameter estimation of ringdown waveforms}

\author{Emanuele Berti} \email{berti@wugrav.wustl.edu}
\affiliation{McDonnell Center for the Space Sciences, Department of Physics, Washington University, St.  Louis,
Missouri 63130, USA}

\author{Jaime Cardoso} \email{jaime.cardoso@inescporto.pt}
\affiliation{INESC Porto, Faculdade de Engenharia, Universidade do Porto, Porto, Portugal}

\author{Vitor Cardoso} \email{vcardoso@phy.olemiss.edu}
\affiliation{Department of Physics and Astronomy, The University of
Mississippi, University, MS 38677-1848, USA \footnote{Also at Centro
de F\'{\i}sica Computacional, Universidade de Coimbra, P-3004-516
Coimbra, Portugal}}

\author{Marco Cavagli\`a} \email{cavaglia@phy.olemiss.edu}
\affiliation{Department of Physics and Astronomy, The University of Mississippi, University, MS 38677-1848,
USA}

\date{\today}

\begin{abstract}
  Using recent results from numerical relativity simulations of non-spinning
  binary black hole mergers we revisit the problem of detecting ringdown
  waveforms and of estimating the source parameters, considering both LISA and
  Earth-based interferometers. We find that Advanced LIGO and EGO could detect
  intermediate-mass black holes of mass up to $\sim 10^3~M_\odot$ out to a
  luminosity distance of a few Gpc. For typical multipolar energy
  distributions, we show that the single-mode ringdown templates presently
  used for ringdown searches in the LIGO data stream can produce a significant
  event loss ($> 10\%$ for all detectors in a large interval of black hole
  masses) and very large parameter estimation errors on the black hole's mass
  and spin. We estimate that more than $\sim 10^6$ templates would be needed
  for a single-stage multi-mode search. Therefore, we recommend a ``two
  stage'' search to save on computational costs: single-mode templates can be
  used for detection, but multi-mode templates or Prony methods should be used
  to estimate parameters once a detection has been made. We update estimates
  of the critical signal-to-noise ratio required to test the hypothesis that
  two or more modes are present in the signal and to resolve their
  frequencies, showing that second-generation Earth-based detectors and LISA
  have the potential to perform no-hair tests.
\end{abstract}


\maketitle


\section{Introduction}

Astronomical observations provide us with a large number of black hole
candidates \cite{Narayan:2005ie}.  Stellar mass candidates (with mass $M\sim
5-20 M_\odot$) are found in X-ray binaries in our galaxy.  Since the discovery
of quasars and other active galactic nuclei, we have growing evidence that
supermassive black holes (SMBHs, with $M\sim 10^5-10^9~M_\odot$) should reside
in the cores of almost all galaxies, including our own.  There is also
mounting observational support to the idea that intermediate-mass black holes
(IMBHs) may fill the gap between these two classes of astrophysical objects.

It is commonly believed that the ultimate test of the ``black hole
hypothesis'' will come from gravitational wave observations (see eg. Lasota's
entertaining account of exotic alternatives to astrophysical black holes
\cite{Lasota:2006jh}, and Hughes' lectures \cite{Hughes:2005wj} for a
relativist's perspective on present and future observational tests of the
black hole hypothesis).  From the point of view of an astrophysicist, black
holes are not particularly interesting: ``black hole candidates'' are
characterized only by their mass and angular momentum, and no compact object
with mass $M\gtrsim 3M_\odot$ has shown any feature that would allow us to
attribute to it any other property other than mass and rotation.  For a
relativist, black holes (being vacuum solutions of the field equations) are
much more exciting: they are unique, ``clean'' probes of the structure of
spacetime in strong-gravity conditions.

An important identifying dynamical feature of a black hole are its
characteristic damped oscillation modes, called quasinormal modes (QNMs)
\cite{kokkotasnollert}. Any compact binary merger leaving behind a black hole
produces a gravitational wave signal that, after the black hole's formation,
can be decomposed as a superposition of exponentially damped sinusoids. By
analogy with the ordinary vibrations of a bell, this signal is known as
``quasinormal ringing'' or ringdown.

The fact that all information is radiated away in the process leading to black
hole formation, so that astrophysical black holes in Einstein's theory are
characterized completely by their mass and angular momentum, is known as the
``no-hair theorem''. For this reason the ringdown signal is very simple: mass
and angular momentum are enough to characterize the black hole's oscillation
spectrum (see Ref.~\cite{bcw} for details), and the detection of ringdown
waves may allow us to identify a black hole and determine its properties. A
measurement of the frequency and damping time of a single QNM can be used, at
least in principle (see below), to determine both the mass and angular
momentum of the hole \cite{echeverria,finn,Flanagan:1997sx,bcw}.  Detection of
more than one QNM would allow us to test consistency with the black hole
hypothesis -- or in more colloquial speaking, to ``test the no-hair theorem''
\cite{Dreyer:2003bv,bcw}.

The detectability of ringdown waves, as well as their use to measure black
hole properties and test relativity, depend mainly on two factors: (i) the
fraction of the black hole's mass radiated in ringdown waves (i.e. the
``ringdown efficiency'' $\epsilon_{\rm rd}$), and (ii) the detector's
sensitivity in the frequency band of interest. In Ref.~\cite{bcw}, some of us
carried out a detailed study of ringdown detectability and no-hair tests with
the planned space-based {\it Laser Interferometer Space Antenna} (LISA). The
ringdown efficiency $\epsilon_{\rm rd}$, the multipolar energy distribution of
the radiation and the dimensionless angular momentum $j$ of the final black
hole were considered as free parameters, to be varied within a certain
reasonable range.  We confirmed and refined earlier estimates by Flanagan and
Hughes \cite{Flanagan:1997sx}, showing that the ringdown signal-to-noise ratio
(SNR) is usually larger than the inspiral SNR for the typical SMBH masses
($M\gtrsim 10^6~M_\odot$) inferred from astronomical observations of nearby
galaxies.

Binary black hole simulations are now carried out by different groups all over
the world. Large-scale simulations of non-spinning unequal-mass binary black
hole mergers were recently used to provide reliable estimates of
$\epsilon_{\rm rd}$, of the final angular momentum $j$ and of the multipolar
energy distribution \cite{Berti:2007fi}.  Ref.~\cite{Berti:2007fi} also showed
that two or more modes are excited to comparable amplitudes in a binary black
hole merger whenever the binary's mass ratio $q\equiv m_2/m_1\neq 1$. The
significant excitation of different multipolar components can be used to
extract useful information on the geometry of the ringing object, and to
perform no-hair tests (in the sense explained above).

In this paper we revisit the analysis of Ref.~\cite{bcw} taking into account
these recent results from numerical relativity, and we extend that study to
include planned and presently operating Earth-based detectors (LIGO, Virgo,
Advanced LIGO and EGO). Many authors have recently stressed the potential of
Earth-based detectors for measuring ringdown waves in the IMBH mass range
\cite{Cutler:2002me,Fregeau:2006yz,Buonanno:2006ui,Pan:2007nw,Baker:2006kr,Ajith:2007qp,Luna:2006gw,Mandel:2007hi,Belczynski:2004gu,Kulczycki:2006ht,Wyithe:2003ju}.
For example, Ref.~\cite{Pan:2007nw} matched an equal-mass merger waveform from
numerical relativity to a Post-Newtonian inspiral and showed that the
resulting SNR for a single LIGO detector peaks at $M\sim 150~M_\odot$, which
is well within the IMBH mass range.  The results presented in this paper
confirm that ringdown waves may be used to provide conclusive evidence for the
existence of IMBHs and (even more importantly) to accurately measure their
parameters.  Initial LIGO (Virgo) may detect IMBHs with $M\lesssim
400~M_\odot$ ($M\lesssim 800~M_\odot$, respectively) out to a luminosity
distance $D_L\sim 100$~Mpc.  Advanced Earth-based interferometers would extend
the luminosity distance at which ringdown events are detectable by a factor
$\sim 10$ for Advanced LIGO, and $\sim 100$ for EGO (see Fig.~\ref{fig:snr}
below). Advanced LIGO and EGO could detect IMBH ringdowns out to cosmological
distances with large SNR, and allow precision measurements of the IMBHs'
properties.

Higher multipolar components of the radiation can be significantly excited
\cite{Berti:2007fi}, and this calls for a critical revision of the
matched-filtering techniques used for ringdown searches. Present ringdown
searches in the LIGO data stream use templates consisting of only one QNM
\cite{goggin} (see also Ref.~\cite{Clark:2007xw} for data analysis methods to
resolve neutron star ringdowns from instrumental glitches).

\begin{figure}[ht]
\epsfig{file=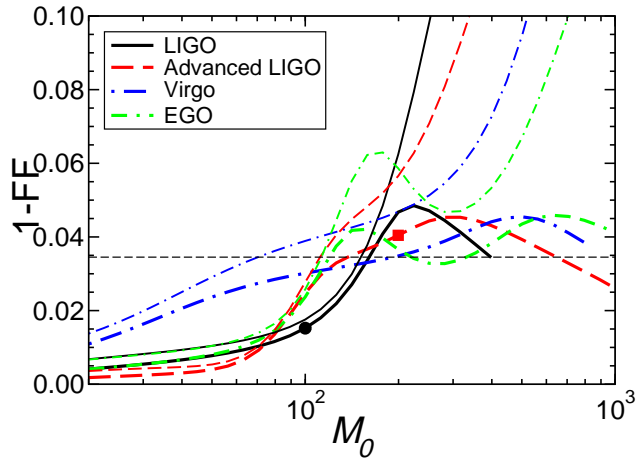,width=7cm,angle=270}
\caption{Owen's minimal match, as defined below Eq.~(\ref{ff}), for different
  Earth-based detectors. For this illustrative calculation we assume that the
  Kerr parameter of the final black hole is $j=0.6$, and that the relative
  amplitude of the second mode is ${\cal A}=0.3$. We also set the phases in
  Eq.~(\ref{twomode}) to be $\phi_1=\phi_2=0$ (thick lines) or $\phi_1=0$,
  $\phi_2=\pi$ (thin lines).  The black circle and the red square mark two
  cases that we study in more detail below: a $M_0=100~M_\odot$ black hole as
  observed by LIGO and a $M_0=200~M_\odot$ black hole as observed by Advanced
  LIGO, respectively.}
\label{fig:ffLIGO}
\end{figure}

In this paper we address this problem and give a preliminary answer to the
following questions:
\begin{itemize}
\item[(1)] Given estimates of the multipolar energy distribution in
  non-spinning binary black hole mergers, how many events would we miss in a
  search with single-mode ringdown templates? The answer is quantified in
  Fig.~\ref{fig:ffLIGO}, where we compute Owen's ``minimal match''
  \cite{Owen:1995tm} as a function of the black hole mass $M_0$ measured in
  the source frame. The event loss is larger than $10\%$ whenever the minimal
  match is larger than $0.035$ (details are provided in the body of the
  paper). From the figure we see that this happens in a significant mass range
  for all Earth-based detectors. We will show below that similar conclusions
  apply also to the planned space-based interferometer LISA, although in a
  completely different mass range.  Furthermore, we will show that single-mode
  templates can produce a large bias in the estimation of the black hole's
  mass and spin. Our conclusions should be rather conservative, because from
  perturbation theory and from numerical relativity results we expect higher
  multipoles to be more excited for black hole binaries with spin and large
  mass ratios.

\item[(2)] How many templates would we need for searches using two-mode
  templates? We estimate that, as compared with single-mode searches, the
  number of templates needed for a two-mode search would increase by roughly
  three orders of magnitude. A more detailed data analysis study (eg. using
  better template placing techniques, along the lines of \cite{nakano}) could
  be useful to reduce computational requirements.

\item[(3)] How strong must a ringdown event be if we want to perform no-hair
  tests or resolve nonlinear contributions \cite{Ioka:2007ak} to the ringdown
  waveform? More precisely: what is the minimum SNR needed to resolve QNMs?
  Below we address different aspects of this problem (frequency, damping time
  and amplitude resolvability require different SNRs). Our results suggest
  that prospects to resolve QNMs are quite optimistic for both LISA and
  second-generation Earth-based detectors.

\end{itemize}

The plan of the paper is as follows. In Sec.~\ref{detectability} we discuss
the potential of Earth-based detectors to measure ringdown waves from
solar-mass black holes and IMBHs.  In Sec.~\ref{sec:evloss} we look at the
event loss and bias in parameter estimation induced by searching for a
two-mode waveform with a single-mode template. In Sec.~\ref{sec:resolvability}
we revise our previous estimates \cite{bcw} of the critical SNR required to
perform no-hair tests. Sec.~\ref{conclusions} contains a brief summary of our
conclusions. Some technical details are presented in the Appendices. In
Appendix \ref{app:numbertemplates} we estimate the number of templates
required to detect multi-mode waveforms, and in Appendix \ref{app:amplres} we
provide a (somewhat optimistic) estimate of the critical SNR required to test
the hypothesis that two modes are present in a ringdown signal.

\section{\label{detectability}Ringdown detectability by Earth-based detectors}

In this Section we study the detectability of ringdown waves by present and
planned Earth-based interferometers: LIGO, Virgo, Advanced LIGO and EGO. Our
analysis complements that of Ref.~\cite{bcw}, where a similar study was
performed for the planned space-based interferometer LISA. Detectability can
be assessed by computing the (maximum) SNR $\rho$ achievable by
matched-filtering:
\be\label{SNRdef}
\rho = \sqrt{(h|h)}\,,
\ee
where the scalar product between two waveforms is defined as
\cite{finn,Apostolatos:1995pj}
\be\label{scalarpr} (h_1|h_2)\equiv
2\int_0^{\infty}\frac{{\tilde h}_1^*(f){\tilde h}_2(f)+{\tilde h}_1(f){\tilde
h}_2^*(f)}{S_h(f)}\,df\,,
\ee
Here
\be\label{ftransform}
{\tilde h}(f)\equiv \int_{-\infty}^{+\infty}e^{2 \pi f t i}h(t) dt\,,
\ee
is the Fourier transform of the waveform $h(t)$, and $S_h(f)$ is the noise
power spectral density (PSD) of the detector. For the initial LIGO and Virgo
PSD we use the analytic approximation of Ref.~\cite{Damour:2000zb}. We assume
that the PSD $S_h(f)=\infty$ for $f<f_s$, where the low-frequency sensitivity
cutoff $f_s=40$~Hz (for initial LIGO) and $f_s=20$~Hz (for Virgo).  For
Advanced LIGO we consider the broadband configuration PSD given in
Ref.~\cite{AdLIGO}, and for the projected PSD of EGO we follow Appendix C of
Ref.~\cite{ego}.
For LISA, we model the PSD (including white-dwarf confusion noise) by the
semianalytic approximation used in \cite{Berti:2004bd,bcw}, with a
low-frequency cutoff $f_s=3\times 10^{-5}$~Hz.

An estimate of the maximum detectable black hole mass for LISA and Earth-based
detectors can be obtained by equating the fundamental $l=m=2$ QNM frequency
(as tabulated and fitted in Ref.~\cite{bcw}) to $f_s$. Since QNM frequencies
scale as $1/M$, this sets a limit on the maximum detectable black hole mass.
The results of this estimate, which are mildly dependent on the black hole's
rotation parameter $j$, are listed in Table~\ref{tab:uppermassbound}.  The
mass range accessible to Earth-based interferometers can increase dramatically
with relatively small (but technically difficult) improvements in the
low-frequency sensitivity threshold. For example, a detector with $f_s$=10~Hz
can detect IMBHs with $M>10^3~M_\odot$ even if they are non-spinning. For the
typical spins that should result from a binary black hole merger ($j\sim
0.6$), a low-frequency sensitivity threshold at $40~$Hz means that initial
LIGO will not be sensitive to QNM ringing from IMBHs with mass $\gtrsim
400~M_\odot$. The detectable mass range increases as the spin of the remnant
black hole gets larger.
\begin{table}[ht]
  \centering \caption{\label{tab:uppermassbound} Upper limit on the detectable
    black hole mass for which the fundamental QNM with $l=m=2$ would be
    detectable by LISA and Earth-based interferometers, for different
    dimensionless spin parameters $j$. A reasonable estimate of $f_s$ is $\sim
    10$~Hz for EGO, and $\sim 20$~Hz for Advanced LIGO (see Ref.~\cite{ego}).}
\begin{tabular}{ccccc}
\hline \hline
  &               $j=0$          & $j=0.6$         &$j=0.7$        &$j=0.98$\\
\hline
{\rm Earth-based}   &$1200 M_{\odot}\left(\f{10~{\rm Hz}}{f_s}\right)$
                    &$1600 M_{\odot}\left(\f{10~{\rm Hz}}{f_s}\right)$
                    &$1720 M_{\odot}\left(\f{10~{\rm Hz}}{f_s}\right)$
                    &$2670 M_{\odot}\left(\f{10~{\rm Hz}}{f_s}\right)$\\
{\rm LISA}          &$4\times 10^8 M_{\odot}$ & $5.3\times 10^8 M_{\odot}$&$5.7\times 10^8 M_{\odot}$&$ 8.9\times 10^8 M_{\odot}$ \\
\hline \hline
\end{tabular}
\end{table}

\begin{figure}[ht!]
\begin{center}
\begin{tabular}{ll}
\psfig{file=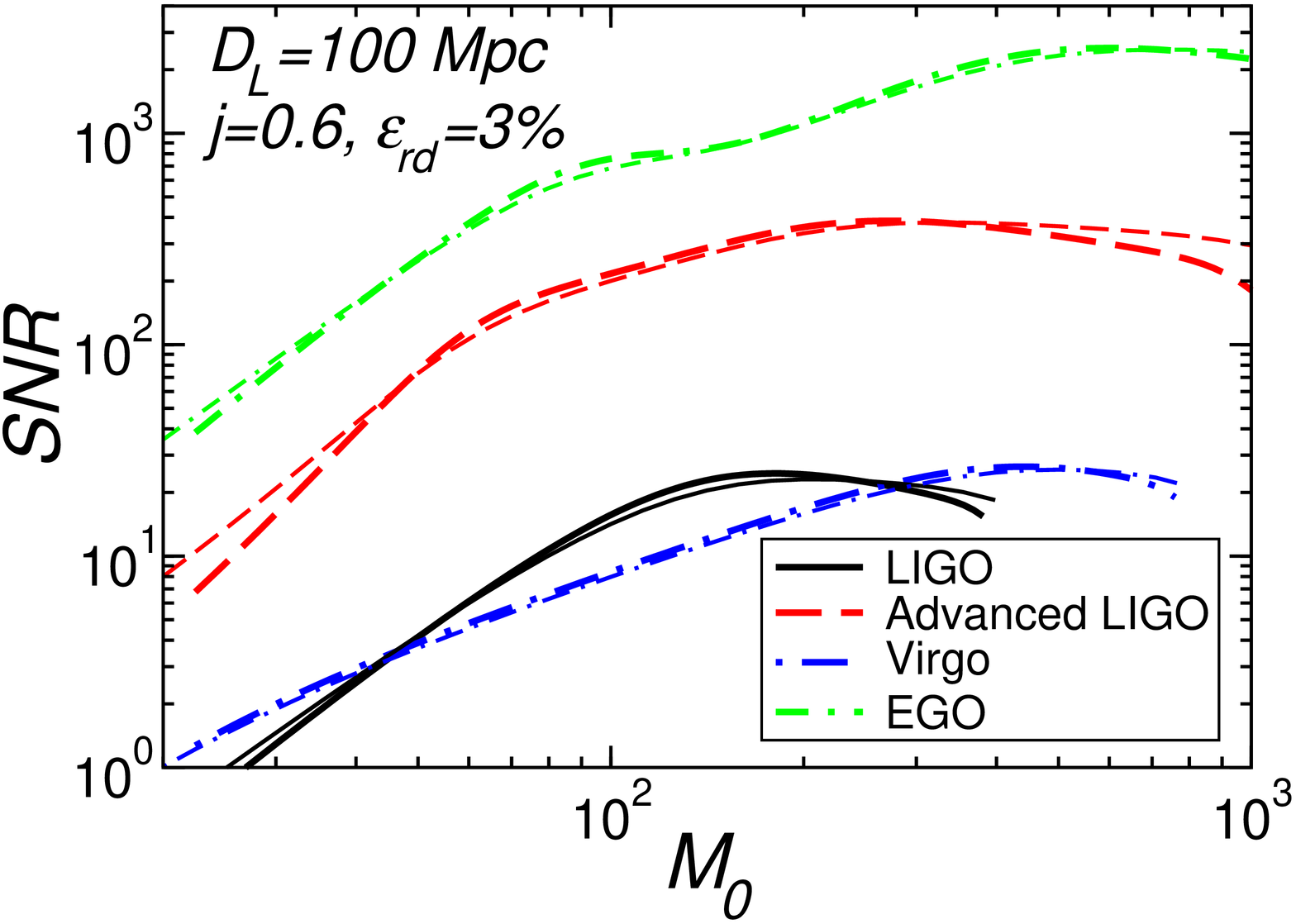,width=7cm,angle=270} &
\psfig{file=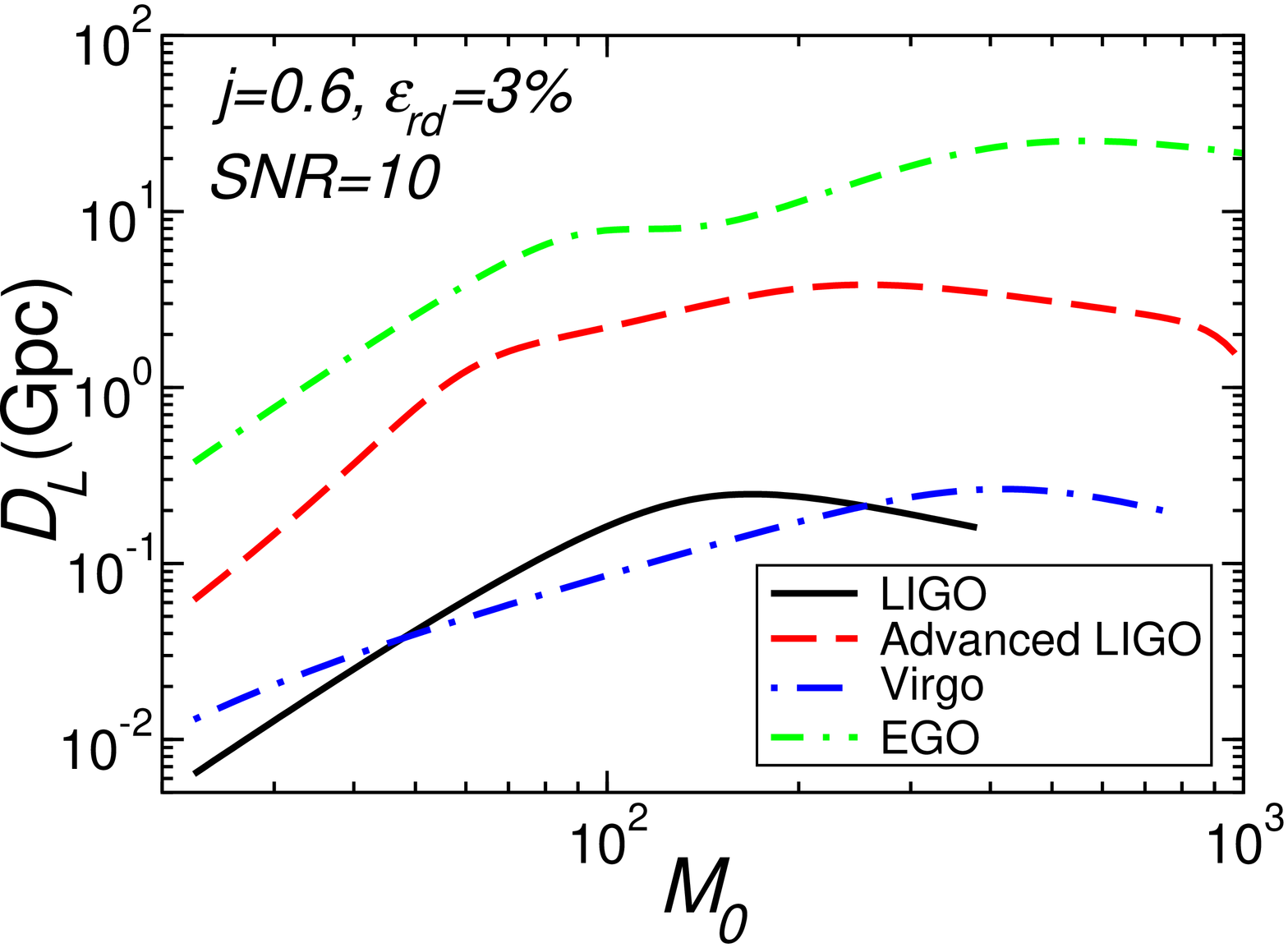,width=7cm,angle=270} \\
\end{tabular}
\end{center}
\caption{Left: ringdown SNR for LIGO, Advanced LIGO, Virgo and EGO at
  $100$~Mpc. Thin lines refer to the EF convention, thick lines to the FH
  convention (see text). Right: luminosity distance (in Mpc) for detectability
  of the ringdown signal with a SNR of 10 (for clarity, here we only show
  results for the FH convention).
}
\label{fig:snr}
\end{figure}

In the left panel of Fig.~\ref{fig:snr} we show the typical sky-averaged SNR
of Earth-based detectors for black hole ringdowns at luminosity distance
$D_L=100$~Mpc as a function of the black hole mass in the source frame $M_0$
(related to the mass $M$ in the detector frame by $M=(1+z)M_0$, where $z$ is
the cosmological redshift). The actual value of the ringdown SNR changes
slightly depending on the way we compute the Fourier transform in
(\ref{ftransform}). There are two common conventions in the literature (see
Ref.~\cite{bcw} for more details). The {\em Echeverria-Finn (EF) convention}
(in the terminology of Ref.~\cite{bcw}) is shown by thin lines in the left
panel.  It assumes that the ringdown waveform $h(t)\sim e^{-t/\tau} \sin(2\pi
f t)$ is zero before some starting time (say $t=0$). Alternatively, we can
adopt the {\em Flanagan-Hughes (FH) doubling prescription} (thick lines in the
figure): we assume that the waveform for $t<0$ is identical to that for $t>0$
except for the sign of $t/\tau$ (i.e., we replace $e^{-t/\tau}$ by
$e^{-|t|/\tau}$) and we divide the resulting SNR by $\sqrt{2}$ to compensate
for the doubling \cite{Flanagan:1997sx}. The left panel of Fig.~\ref{fig:snr}
shows that adopting one or the other convention does not change the results
significantly. In the remainder of the paper, we will adopt the EF convention,
since it is generally more straightforward to implement in matched-filtering
applications.

The right panel of Fig.~\ref{fig:snr} shows the luminosity distance out to
which ringdown would be detectable by each detector with a SNR of 10 (assuming
a flat, $\Lambda$-dominated cosmology, in accordance with the latest
observational data).
%
%
In the Figure we use typical values from simulations of non-spinning binary
black hole mergers \cite{Berti:2007fi}, assuming that the final black hole has
angular momentum $j=0.6$ and that the ringdown efficiency is $\epsilon_{\rm
  rd}\simeq 3\%$. The results would not change much had we made different
assumptions. In particular, the SNR scales with efficiency as
$\rho\sim\sqrt{\epsilon_{\rm rd}}$. If the merging black hole binary has mass
ratio $q\equiv m_2/m_1< 1$, the ringdown SNR would decrease (to a very good
approximation) by a factor $4q/(1+q)^2$ compared to the equal-mass case
\cite{Berti:2007fi}.  Fig.~\ref{fig:snr} can be compared with Figs.~7 and 8 in
Ref.~\cite{bcw}, showing the inspiral and ringdown SNR of LISA at different
values of the cosmological redshift (for clarity, in Fig.~\ref{fig:snr} we do
not show the inspiral SNR). We also note that Figure 2 assumes matched
filtering with one mode only, and that (following results from numerical
relativity simulations) we assumed this mode to be the fundamental mode with
$l=m=2$.  Inclusion of higher multipoles, the main topic of this paper, would
only increase the sky-averaged SNR by small factors of order unity, but (as we
will show) it may have important implications for a matched-filtering
detection of the signal.

The left panel of Fig.~\ref{fig:snr} is in good agreement with the results for
LIGO and Advanced LIGO shown in Fig.~2 of Ref.~\cite{Fregeau:2006yz}. Our
results for Advanced LIGO are slightly different because we use a more
accurate model for the PSD. For Virgo, our results agree with those in
Ref.~\cite{Kulczycki:2006ht}.  The right panel clarifies the potential of
present and future interferometric detectors to detect ringdown waves.  At the
present sensitivity, LIGO could detect waves from IMBHs of mass $M\sim 10^2
M_\odot$ out to a distance of about $100~$Mpc. Virgo spans roughly the same
distance range, but it should be able to detect larger IMBH masses when the
low-frequency design sensitivity is met.  Advanced LIGO will extend the
distance range out to a few Gpc (a factor $\sim 10$) and it should be able to
detect IMBHs of mass $10^2~M_\odot<M_0<10^3~M_\odot$.  EGO would be a
remarkable tool to detect IMBHs of mass as large as $\sim 10^3~M_\odot$ out to
distances of $\sim 10~$Gpc and larger, and to measure their properties with
remarkable precision.

\section{\label{sec:evloss} Event loss and bias in parameter estimation using single-mode templates}

A common choice to search for signals of known form in noisy data is
matched-filtering. Matched-filtering works by cross-correlating the signal
with a set of theoretical templates\footnote{For alternative techniques
  especially designed to extract damped sinusoidal signals from noise, see
  Ref.~\cite{bcgs} and references therein.}. Current searches for ringdown
signals in the LIGO data stream \cite{goggin} and in resonant bar detectors
\cite{Pai} make use of the simplest theoretical model: a single damped
sinusoid. Here we try to assess the performance of such a template to detect a
{\it superposition} of damped sinusoids. More precisely, we estimate the
number of events we would miss by using single-mode templates to detect
multi-mode signals, and we try to quantify the bias in measured parameters
induced by the use of such templates.

The optimal way of addressing the performance of single-mode ringdown
templates would be to use our present best guess for the ``true'' waveform
emitted by a binary black hole merger: the wavetrain obtained by stitching
together the Post-Newtonian predictions for the inspiral and the best
available numerical relativity waveforms. Finding the optimal way to perform
this stitching is in itself a difficult problem, now actively pursued by many
research groups \cite{Buonanno:2006ui,Pan:2007nw,Baker:2006ha,Berti:2007fi}.
Our purpose here is to make some general points of principle, exploring
semi-quantitatively the typical event loss and parameter bias induced by
single-mode templates.  For this reason, and to keep the analysis as simple
and general as possible, our ``true'' waveform will be taken to consist of a
two-mode QNM superposition.  Therefore we assume that the following two-mode
signal is impinging on the detector, starting at time $t=0$:
\begin{subequations}
\label{twomode}
\beq
h(t)
&=&
{\cal A}_1
e^{-(\pi f_{1}/Q_{1}) t}\sin\left (2\pi f_{1}t-\phi_1\right )+
{\cal A}_2
e^{-(\pi f_{2}/Q_{2}) t}\sin\left (2\pi f_{2}t-\phi_2\right )\\
&=&
{\cal A}_1\left[
e^{-(\pi f_{1}/Q_{1}) t}\sin\left (2\pi f_{1}t-\phi_1\right )+
{\cal A}
e^{-(\pi f_{2}/Q_{2}) t}\sin\left (2\pi f_{2}t-\phi_2\right )\right]\,.
\eeq
\end{subequations}
Here $f_i$ ($i=1,~2$) denotes the oscillation frequency of each QNM,
$Q_i\equiv \pi f_i \tau_i$ is the quality factor of the oscillation, and
${\cal A}_i$ is the oscillation amplitude of mode $i$. For reasons that will
become apparent in the following, in the second line we found it convenient to
express the signal in terms of the relative amplitude of the two modes ${\cal
  A}\equiv {\cal A}_2/{\cal A}_1$.  The Fourier transform of
Eq.~(\ref{twomode}) in the EF convention reads:
\beq\label{FTsignal}
{\tilde h}(f)=
{\cal A}_1
\frac{Q_1 \left [2f_1Q_1\cos\phi_1-(f_1-2ifQ_1)\sin\phi_1\right ]}
{\pi \left [f_{1}^2-4iff_{1}Q_{1}+4(-f^2+f_{1}^2)Q_{1}^2\right ]}+
(1\to 2)
\,.
\eeq

Matched-filtering works by cross-correlating the detector's output with a set
of templates. Consider a one-mode bank of templates such as those presently
used in ringdown searches \cite{goggin}:
\beq \label{template}
T\{{\vec \lambda}\}&=&
e^{-\pi f_T/Q_T t}\sin[2\pi f_T (t-t_0)-\phi_T]\,,\qquad
{\rm for \,\,} t\geq t_0\,.
\eeq
where $\vec\lambda$ is an ($N$-dimensional) vector of template parameters, and
$T\{{\vec \lambda}\}=0$ for $t<t_0$. In this particular case,
$\vec\lambda=(f_T\,,Q_T\,,\phi_T)$ and $N=3$ (in principle we could use $t_0$
as an additional parameter, but we verified by explicit calculations that
setting $t_0=0$ doesn't sensibly affect the fitting factor and the estimated
values of the other parameters). The Fourier transform of this template is
similar to those in Eq.~(\ref{FTsignal}), with an additional exponential
factor depending on the starting time $t_0$:
\be
{\tilde T}(f)=
\frac{Q_T \left [2f_TQ_T\cos\phi_T-(f_T-2ifQ_T)\sin\phi_T\right ]}
{\pi \left( f_T^2-4iff_TQ_T+4(-f^2+f_T^2) Q_{T}^2 \right ) }
\exp\left[{\f{-\pi(f_T-2ifQ_T)t_0}{Q_T}}\right] \,.
\ee

The performance of the template (\ref{template}) to detect the two-mode
waveform (\ref{twomode}) can be computed with the help of the fitting factor
(FF) first introduced by Apostolatos \cite{Apostolatos:1995pj}:
\be
{\rm FF}=\max_{\vec \lambda}
\frac{(T\{{\vec \lambda}\}|h)}
{\sqrt{(T\{{\vec \lambda}\}|T\{{\vec \lambda}\})(h|h)}}
\label{ff}\,.
\ee
where the scalar product between two waveforms has been defined in
Eq.~(\ref{scalarpr}). An equivalent quantity often used in the literature is
Owen's ``minimal match'', defined as $(1-{\rm FF})$ \cite{Owen:1995tm}.
From the definition, Eq.~(\ref{ff}), it is clear that an overall normalization
constant (say, ${\cal A}_1$) does not affect calculations of the FF: this is
the reason for introducing the relative amplitude ${\cal A}$ defined in
(\ref{twomode}).  The FF measures the degradation of the SNR due to cross
correlating an arbitrary signal $h(t)$, such as the two-mode signal
(\ref{twomode}), with all filters $T\{\vec \lambda\}$ in the template bank.
The effective SNR that can be obtained by matched-filtering is
\be\label{rhomf}
\rho_{\rm MF}=\max_{\vec \lambda}
\frac{(T\{{\vec \lambda}\}|h)}
{\sqrt{(T\{{\vec \lambda}\}|T\{{\vec \lambda}\})}}=
{\rm FF}\times \rho\,,
\ee
where $\rho$ has been defined in Eq.~(\ref{SNRdef}). For gravitational wave
detection the SNR is proportional to the inverse of the luminosity distance to
the source, while the event rate (scaling with the accessible volume) is
proportional to the cube of this distance.  Therefore, given the FF, we can
compute the event loss as
\be {\rm event \, loss}=1-{\rm FF}^3\,. \ee
For detection purposes one usually requires ${\rm FF}\gtrsim 0.965$,
corresponding to a loss of less than $10\%$ of the events that could be
potentially be detected by a ``perfect'' filter.

We compute the integrals in (\ref{ff}) and (\ref{rhomf}) using a
Gauss-Legendre spectral integrator, and we perform the maximization by a
Fortran implementation of the Nelder-Mead downhill simplex method
\cite{numrec}. In actual searches, the set of all templates forms a grid that
should cover the $N$-dimensional parameter space.  The point on this grid for
which the FF is maximum singles out the template that best matches the actual
waveform impinging on the detector. If the FF is above some pre-determined
threshold, eg. ${\rm FF}\gtrsim 0.965$, we say that we have a detection.  An
estimate of the number of templates required to detect multi-mode waveforms is
given in Appendix \ref{app:numbertemplates}.

The detectability of a multi-mode signal by a single-mode template depends,
among other things, on the relative amplitude of the subdominant modes in the
``real'' signal. In our simplified signal (\ref{twomode}) we denoted this
relative amplitude by ${\cal A}$. Below we discuss how recent numerical
simulations of binary black hole mergers can be used to provide such an
estimate.

\subsection*{Estimates of the relative mode amplitude from numerical relativity}

Recent results from numerical simulations of the merger of non-spinning,
unequal mass black hole binaries support the expectation that the $l=m=2$ mode
should dominate the ringdown signal \cite{Berti:2007fi}. However, they also
reveal that other modes (especially $l=m=3$ and $l=m=4$) are significantly
excited.  Whenever the mass ratio $q\gtrsim 2$, the $l=m=3$ mode is excited to
roughly one third of the amplitude of the $l=m=2$ mode. We expect this
estimate of the relative excitation to be conservative: perturbative results
and numerical simulations indicate that higher multipoles should be more
excited as the the binary's mass ratio grows, or when the black holes in the
binary are spinning.

Merger simulations show that the projection of the Weyl scalar $\Psi_4$ onto
spin-weighted spherical harmonics ${_{-2}}Y_{lm}(\theta\,,\phi)$ has a
circular polarization pattern (see Appendix D of Ref.~\cite{Berti:2007fi}). In
the ringdown regime the Weyl scalar $\Psi_4$ can further be decomposed as a
QNM sum, and to a good approximation we can write:
\be \Psi_4=\frac{1}{r}\sum _{lm}\omega_r^2 {\cal A}_{l|m|}e^{-t/\tau}
{_{-2}}Y_{lm}(\theta\,,0)\times
\left[\sin\left (\chi-|m|\phi\right )+i \sin\left (\chi-|m|\phi+\pi/2 \right ) \right ] \,.\ee
For simplicity, in this expansion we are considering only the least damped,
fundamental mode for each $(l\,,m)$ pair \cite{bcw}. We also approximate
spin-weighted spheroidal harmonics by spin-weighted spherical harmonics, which
introduces an error of order $\sim 1\%$ or less \cite{Berti:2005gp}.
The variable $\chi\equiv \omega_r t+\varphi_0$, with $\varphi_0$ a constant,
and it is implicitly assumed that all frequencies and damping times depend on
the angular numbers $(l\,,m)$. Using the equatorial symmetry of the system,
the waveform can be rewritten as
\be \Psi_4=\frac{1}{r}\sum _{l\,,m>0} \omega_r^2{\cal A}_{l|m|}e^{-t/\tau}\times  \left[Y_{lm}^+\sin\left
(\chi-m\phi\right )+i Y_{lm}^{\times}\sin\left (\chi-m\phi+\pi/2 \right ) \right ]\,, \ee
where we have defined the following useful quantities:
\begin{subequations}
\beq Y_{lm}^+&\equiv& {_{-2}}Y_{lm}(\theta\,,0)+(-1)^l{_{-2}}Y_{l-m}(\theta\,,0)\,,\\
Y_{lm}^{\times}&\equiv&
{_{-2}}Y_{lm}(\theta\,,0)-(-1)^l{_{-2}}Y_{l-m}(\theta\,,0)\,. \eeq
\end{subequations}
%
%
Recalling that $\Psi_4 = \ddot{h}_+-i\ddot{h}_{\times}$ and using the
large-$Q$ limit, which is usually a good approximation \cite{bcw}, we get
\begin{subequations}
\beq
h_{+}&=&-\frac{1}{r} \sum _{l\,,m>0}{\cal A}_{l|m|}e^{-t/\tau}Y_{lm}^+\sin\left
(\chi-m\phi\right ) \,,\\
h_{\times}&=&\frac{1}{r}\sum _{l\,,m>0} {\cal A}_{l|m|}e^{-t/\tau} Y_{lm}^{\times}\sin\left
(\chi-m\phi+\pi/2
\right )\,.
\eeq
\end{subequations}
The gravitational wave strain at the detector is given by
\beq h =
h_{+} F_+(\theta_S,\phi_S,\psi_S)+
h_{\times} F_\times(\theta_S,\phi_S,\psi_S) \,, \label{detectwave}
\eeq
where $F_{+,\times}(\theta_S,\phi_S,\psi_S)$ are pattern functions depending
on the orientation of the detector and on the direction of the source, whose
expressions can be found (for example) in Ref.~\cite{thorne}. In the
following, to simplify the notation, we will drop the functional dependence of
$Y_{lm}^{+,\times}$ and $F_{+,\times}$ on the angles.

%
%
Let us consider, for simplicity, a two-mode situation. Guided by numerical
results for the merger phase (see below) we can assume that the dominant modes
are $l=m=2$ and $l=m=3$. Then we get
\begin{widetext}
\beq\label{hangles}
h(t)&=&
\frac{{\cal A}_{22}}{r}
\left\{
e^{-t/\tau_{22}}
\left [
\sin\left (\omega_r^{22}t+\varphi_{22}-2\phi\right )
Y_{22}^+ F_+
+\sin\left(\omega_r^{22}t+\varphi_{22}-2\phi+\pi/2 \right )
Y_{22}^{\times}F_{\times} \right
]-\, \right. \nonumber\\
& &
\left.
\f{{\cal A}_{33}}{{\cal A}_{22}}
e^{-t/\tau_{33}}
\left [
\sin\left(\omega_r^{33}t+\varphi_{33}-3\phi\right )
Y_{33}^+F_+ +
\sin\left (\omega_r^{33}t+\varphi_{33}-3\phi+\pi/2 \right)
Y_{33}^{\times}F_{\times}
\right ]
\right\}\,.
\label{response}
\eeq
\end{widetext}

The relative magnitude of different multipolar components depends on the
factor ${\cal A}={\cal A}_{33}/{\cal A}_{22}$ (discussed below), but it is
also a function of the angles $(\theta,\phi)$: that is, it depends on the
orientation of the black hole's spin. Let us discuss the influence of these
two factors, in turn.

\begin{table}[ht]
 \centering \caption{\label{tab:relamps} Relative amplitudes of different
   multipoles, during the ringdown phase.}
\begin{tabular}{ccccc}
\hline
\hline \multicolumn{1}{c}{} & \multicolumn{2}{c}{EMOP}& \multicolumn{2}{c}{Peak}\\%
\hline
$q$ &${\cal A}_{33}/{\cal A}_{22}$ & ${\cal A}_{44}/{\cal A}_{22}$ &
${\cal A}_{33}/{\cal A}_{22}$ & ${\cal A}_{44}/{\cal A}_{22}$    \\
1  & 0.00    & 0.05 &0.00&0.06\\
1.5& 0.09    & 0.05 &0.12&0.06\\
2.0& 0.15    & 0.05 &0.19&0.06\\
2.5& 0.19    & 0.06 &0.24&0.08\\
3.0& 0.20    & 0.06 &0.28&0.09\\
3.5& 0.21    & 0.07 &0.32&0.10\\
4.0& 0.23    & 0.08 &0.35&0.12\\
\hline \hline
\end{tabular}
\end{table}
To start with, in Table \ref{tab:relamps} we show two different estimates of
the relative amplitudes ${\cal A}_{33}/{\cal A}_{22}$ and ${\cal A}_{44}/{\cal
  A}_{22}$ from numerical simulations of unequal-mass black hole binaries.
The first estimate uses the energy maximized orthogonal projection (EMOP)
criterion. The idea is to slide a ringdown template of the form
(\ref{template}) along the numerical waveform, and define the starting time of
the ringdown as the time maximizing the ``energy content'' of the ringdown
wave, which can be defined using a suitable scalar product. Given this
starting time, we can also compute the energy radiated in a given multipolar
component of the ringdown waveform according to the EMOP criterion, $E^{\rm
  EMOP}_{lm}$ (see Ref.~\cite{Berti:2007fi} for details). The radiated energy
$E^{\rm EMOP}_{lm} \propto \left({\cal A}^{\rm EMOP}_{lm}\right)^2 Q_{lm}
\omega_{lm}$, where ${\cal A}^{\rm EMOP}_{lm}$ is the amplitude of the
physical waveform $h$, $\omega_{lm}$ the vibration frequency of the
fundamental mode and $Q_{lm}$ its quality factor \cite{bcw}. Then our EMOP
estimate of the relative amplitude of different multipolar components is
\be
\f{{\cal A}^{\rm EMOP}_{lm}}{{\cal A}^{\rm EMOP}_{l'm'}}=
\left[
\frac{E_{lm}Q_{l'm'}\omega_{l'm'}}{E_{l'm'}Q_{lm}\omega_{lm}}
\right]^{1/2}\,.
\ee

To bracket uncertainties, a second estimate can be obtained by simply
computing the ratio of the moduli of the waveforms $|h^{\rm
  peak}_{lm}|/|h^{\rm peak}_{l'm'}|$, where a superscript ``peak'' means that
we evaluate the modulus of the waveform's amplitude at the maximum (see
Ref.~\cite{Berti:2007fi}). We call this the ``peak estimate''.

The amplitude ratios depend on the modes being considered and on the binary's
mass ratio $q$, and their functional dependence on $q$ can be deduced from the
leading-order Post-Newtonian quasicircular approximation discussed in
Ref.~\cite{Berti:2007fi}. We find that the data in Table \ref{tab:relamps} are
well approximated by the following fitting relations:

\begin{subequations}
\beq\label{Aq}
\f{{\cal A}_{33}}{{\cal A}_{22}}
&\simeq& k_1 (1-1/q)\,,\\
\f{{\cal A}_{44}}{{\cal A}_{22}}
&\simeq& k_2 +k_3  \f{q^2}{(1+q)^2}\,
\eeq
\end{subequations}
where the values of the fitting constants depend on the estimation method we
use. If we use the EMOP criterion we get
$k_1^{\rm EMOP}=0.303$,
$k_2^{\rm EMOP}=-0.0134$,
$k_3^{\rm EMOP}=0.1400$.
If instead we use the peak estimate, the fitting coefficients are
$k_1^{\rm peak}=0.431$,
$k_2^{\rm peak}=-0.0670$,
$k_3^{\rm peak}=0.2843$.

\begin{figure}[ht]
\begin{center}
\epsfig{file=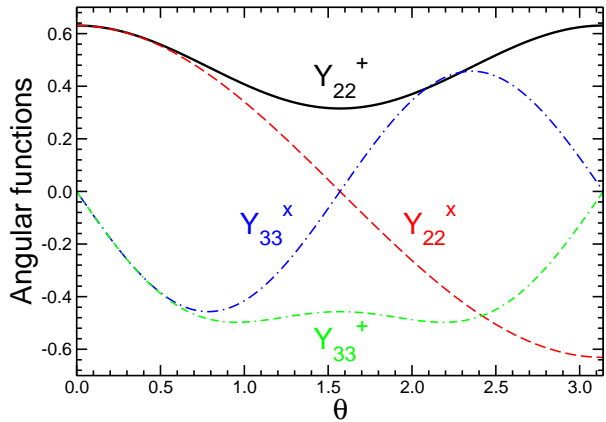,width=7cm,angle=270}
\end{center}
\caption{$\theta$-dependent angular functions in Eq.~(\ref{response}).}
\label{fig:sphericalharmonics}
\end{figure}

Let us now turn to the angular dependence of different multipolar components.
From Eq.~(\ref{hangles}) we see that the relative amplitude of the modes is a
complicated function of the sky position and orientation of the source,
depending on products of the angular functions $F_{+,\times}$ and
$Y_{lm}^{+,\times}$. A possible way to determine the influence of these angles
on the relative amplitude of the modes would be to perform Monte Carlo
simulations, assuming random distributions for the angles. This is beyond the
scope of this paper. Some insight can be obtained by plotting the
$\theta$-dependent combinations appearing in Eq.~(\ref{hangles})
(Fig.~\ref{fig:sphericalharmonics}). For simplicity, let us consider the
following two cases:

\begin{itemize}

\item[(i)] The angles $(\theta_S,\phi_S,\psi_S)$ are such that $F_\times=0$.
  Then the relative amplitude ${\cal A}_2/{\cal A}_1$ depends on the product
  of two factors: $({\cal A}_{33}/{\cal A}_{22})\times (Y_{33}^+/Y_{22}^+)$.
  In this case, from Fig.~\ref{fig:sphericalharmonics} we see that the
  ``plus'' component of the subdominant ($l=m=3$) multipole is never
  significantly enhanced with respect to the dominant multipole, because
  $|Y_{33}^+/Y_{22}^+|\lesssim 1$ for all values of $\theta$.

\item[(ii)] The angles $(\theta_S,\phi_S,\psi_S)$ are such that $F_+=0$. In
  this case the angular factor $(Y_{33}^\times/Y_{22}^\times)$ can blow up at
  the equator $\theta=\pi/2$, so that the ``subdominant'' $l=m=3$ component
  will actually dominate the waveform for an observer located in this
  direction.

\end{itemize}

In conclusion, the relative amplitude ${\cal A}\equiv {\cal A}_2/{\cal A}_1$
is at most $\sim 1/3$ for non-spinning mergers with moderate mass ratios.
Subdominant components can be amplified by the angular dependence of the
radiation, but the likelihood of such an amplification should be quantified by
a more detailed analysis.  If the ratio of angular functions is of order
unity, a relative amplitude ${\cal A}\simeq 0.3$ can be thought of as a
conservative estimate. From both point particle results and present-day
simulations of spinning binaries in numerical relativity, we expect higher
multipoles to be more excited for initially spinning black holes and large
mass ratios. Below we illustrate the typical effect of subdominant multipoles
on matched-filtering assuming ${\cal A}=0.3$.

\subsection*{Fitting factors and event loss due to single-mode templates}

In our calculations of the FF we will assume that the angular dependence of
the waveform (\ref{response}) is such that the signal can be simplified to the
form of Eq.~(\ref{twomode}). This assumption is not general enough. Our
discussion above shows that it will only be valid for specific source
locations in the sky, or for specific orientations of the detector: for
example, the signal simplifies to Eq.~(\ref{twomode}) when $F_+=0$ or
$F_\times=0$, as long as we consider an observer located at some constant
$\theta$ and we appropriately define the azimuthal angles $(\phi_1\,,\phi_2)$.
A good strategy to address the general case could make use of Monte Carlo
simulations. However, the simplified waveform (\ref{twomode}) captures many of
the important features we wish to address in this paper.  Based on our
discussion of the relative multipolar excitation we assume that mode ``1'' is
the fundamental QNM with $l=m=2$, and that mode ``2'' is the fundamental QNM
with $l=m=3$.  We set the relative amplitude of the two modes ${\cal A}=0.3$,
and to compute the QNM frequencies and quality factors we consider a
dimensionless Kerr parameter $j=0.6$ (a typical value for the end-product of
unequal-mass, non-spinning binary black hole mergers).

Fig.~\ref{fig:ffLIGO} shows Owen's ``minimal match'' $(1-{\rm FF})$
\cite{Owen:1995tm} for this choice of parameters. The calculation is performed
for different Earth-based detectors (LIGO, Virgo, Advanced LIGO and EGO) and
the minimal match is computed as a function of the black hole's mass in the
source rest frame, $M_0$.  Thick lines assume that $\phi_1=\phi_2=0$: roughly
speaking, this means that the subdominant mode is ``in phase'' with the
dominant multipole. Thin lines assume $\phi_1=0$ and $\phi_2=\pi$, a rough way
to simulate ``dephased'' signals.  Intuitively we would expect a single-mode
template to be a worse match for dephased signals.  This expectation is
confirmed by the fact that the minimal match is usually larger when
$\phi_2=\pi$ (see below for a more detailed analysis). Values above the
horizontal dashed line, corresponding to a ${\rm FF}=0.965$, correspond to an
event loss larger than $10\%$. This illustrative calculation shows that, for
the typical relative amplitudes expected from numerical relativity, {\it
  single-mode templates can produce a significant event loss in Earth-based
  detectors when the black hole mass $M\gtrsim 10^2~M_\odot$}.  This event
loss can be fatal for detection of signals from very large mass IMBHs
($M_0\gtrsim 200~M_\odot$), especially when the $l=m=2$ and $l=m=3$ components
are not in phase. A more detailed discussion of the dependence of the FF on
the phases $(\phi_1\,,\phi_2$) can be found below.

\begin{figure}[ht]
\begin{tabular}{ll}
\epsfig{file=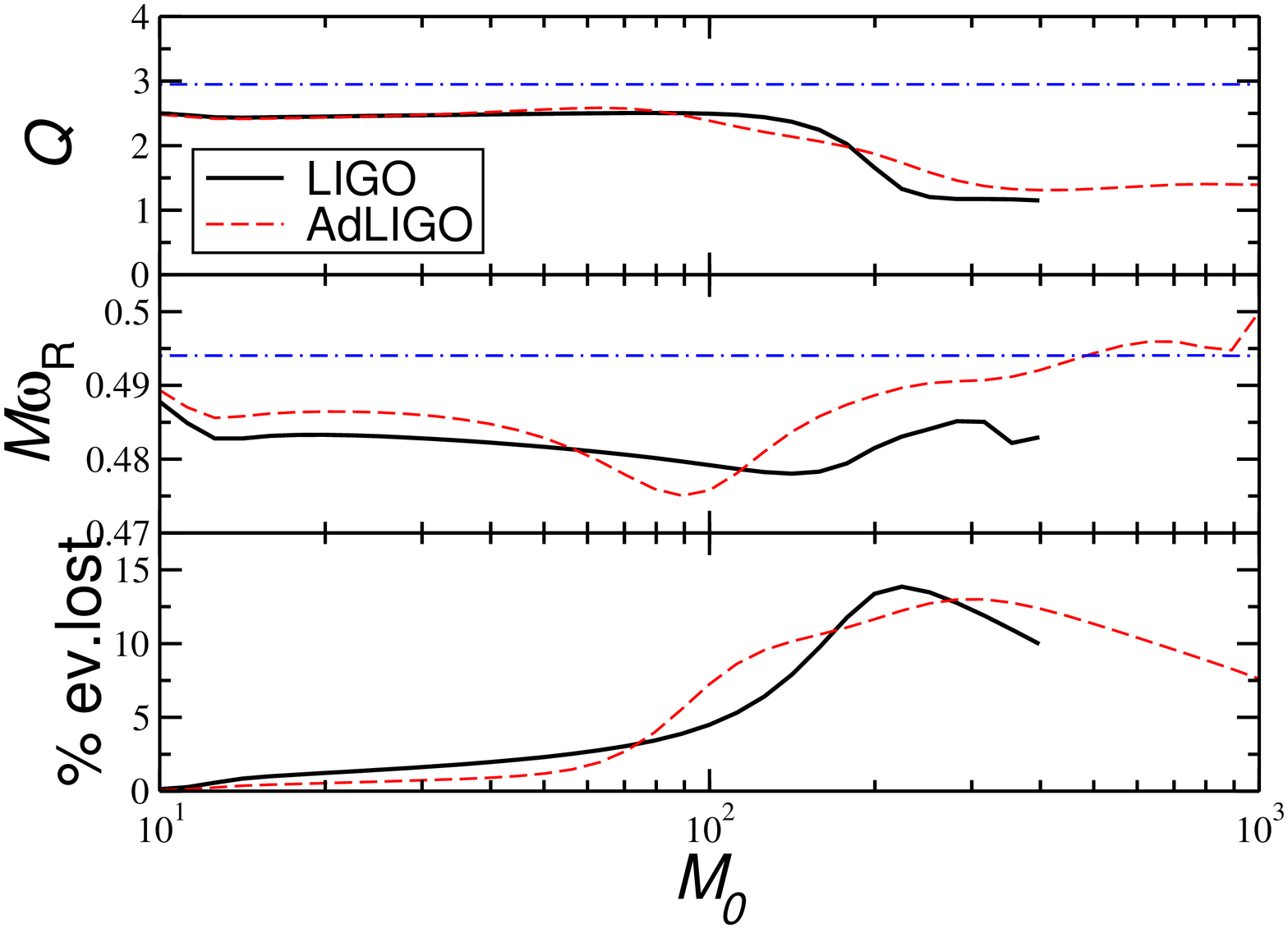,width=7cm,angle=270} &
\epsfig{file=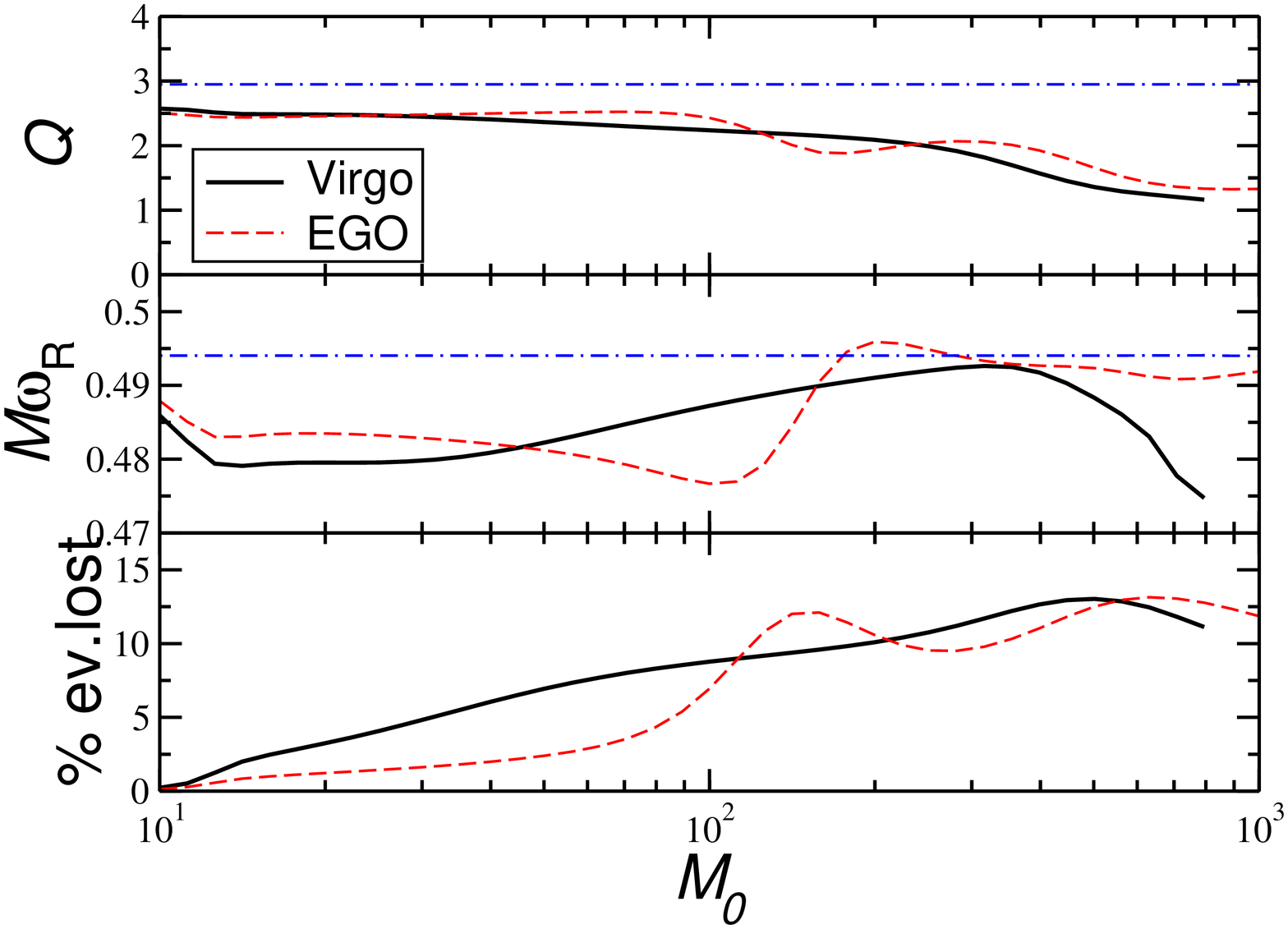,width=7cm,angle=270} \\
\end{tabular}
\caption{Quality factor (top panel) and frequency (middle panel) maximizing
  the FF, and corresponding event loss (bottom panel), for Earth-based
  detectors. Solid and dashed lines are the template's frequency and quality
  factor maximizing the FF, and the corresponding event loss.  Horizontal
  lines show the frequency and quality factor of the fundamental mode with
  $l=m=2$.}
\label{fig:LIGO}
\end{figure}

\begin{figure}[ht]
\begin{tabular}{ll}
\epsfig{file=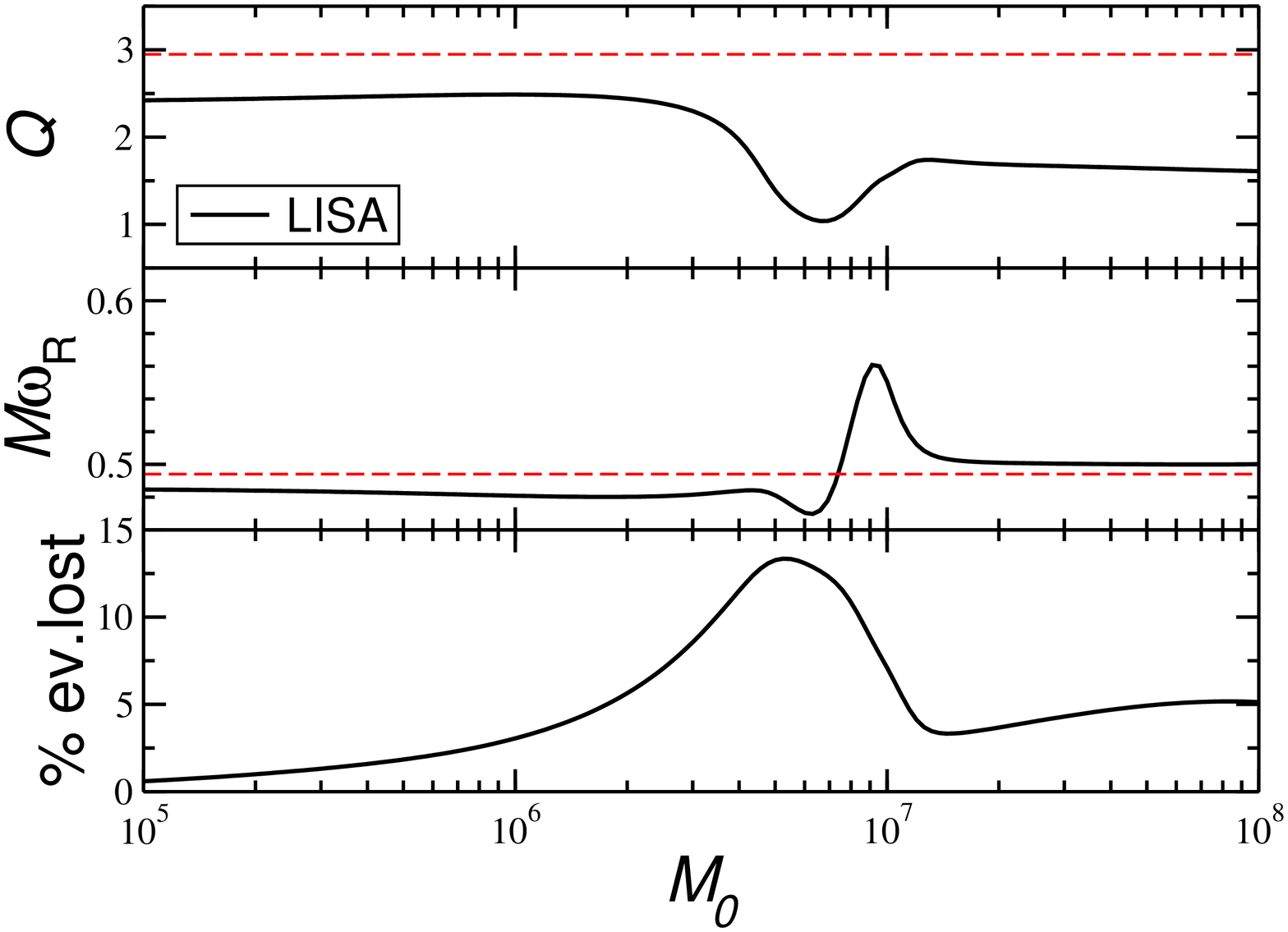,width=7cm,angle=270} &
\epsfig{file=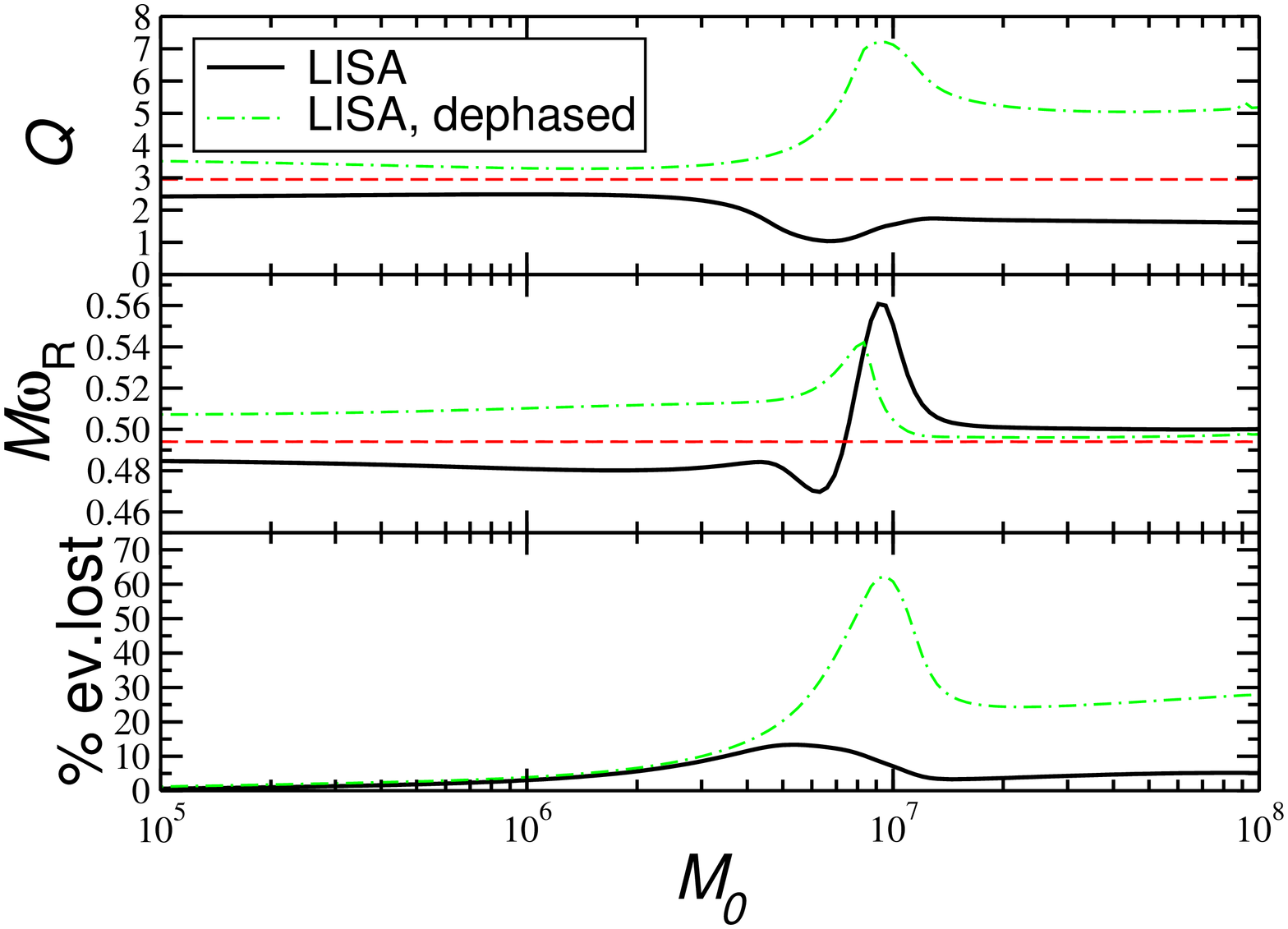,width=7cm,angle=270}
\end{tabular}
\caption{Same as Fig.~\ref{fig:LIGO} for LISA. In the left panel we choose
  $\phi_1=\phi_2=0$, as for Earth-based detectors. In the right panel we also
  consider a ``dephased'' QNM superposition with $\phi_1=0$, $\phi_2=\pi$
  (green, dot-dashed lines).}
\label{fig:LISA-dephase}
\end{figure}

A low value of the FF is usually accompanied by a large bias in parameter
estimation. This is illustrated quantitatively in Figs.~\ref{fig:LIGO} and
\ref{fig:LISA-dephase}, where we plot the quality factor and (dimensionless)
frequency maximizing the FF, and the corresponding event loss, for different
detectors. In Fig.~\ref{fig:LIGO} we consider Earth-based detectors of first
and second generation. To be conservative, we perform the calculation in the
``optimistic'' case when the $l=m=2$ and $l=m=3$ modes start in phase
($\phi_1=\phi_2=0$, ${\cal A}=0.3$).

The QNM frequency of the dominant mode (horizontal dash-dotted line) is
usually estimated with very good accuracy by a single-mode filter, except for
very large values of the mass ($M\gtrsim 500~M_\odot$ for LIGO and Virgo, and
$M\gtrsim 10^3~M_\odot$ for second-generation detectors).  Unfortunately, even
when the FF is very high the single-mode filter produces a large bias in the
quality factor of the dominant mode. For all Earth-based detectors we
consider, this error is $\gtrsim 20\%$ even for low values of the mass
($M_0<100~M_\odot$), when the event loss is quite low and the filter works
well for the purpose of detection.  The bias on the quality factor gets even
worse when we allow for a possible dephasing of the subdominant multipole (see
Fig.~\ref{fig:SNRQcontours} below).  Being dimensionless, the quality factor
of a QNM depends only on the dimensionless angular momentum $j$ of the black
hole \cite{bcw}.  Therefore, a large bias in the quality factor means that
single-mode templates cannot be used for accurate measurements of the black
hole's angular momentum, contrary to early claims in the literature (see e.g.
Ref.~\cite{echeverria}).  Of course, this does not mean that such measurements
are not possible. Single-mode templates are useful for detection (at least for
small black hole masses), but multi-mode templates will be necessary if we
want to perform precision measurements of a black hole's properties using
ringdown waves.

Fig.~\ref{fig:LISA-dephase} shows that these remarks remain valid even for
LISA, when the SNR is much larger and the detectable QNM frequencies much
lower. In the plot we consider a source at $D_L=1~$Gpc ($z\sim 0.2$), for
which the SNR can be $\gtrsim 10^3$ \cite{bcw}. The left panel shows the
``optimistic'' case when the first and second QNM signals are in phase. Even
in this optimal situation the event loss can be as large as $\sim 15\%$ for
masses $M_0\sim 5\times 10^6~M_\odot$, roughly the measured mass of the SMBH
at the center of our own Galaxy.  When the two QNM signals are dephased
(green, dot-dashed line in the right panel) the event loss can be larger then
$60\%$ for masses $M_0\sim 10^7~M_\odot$. Notice also that for $M_0\lesssim
10^7~M_\odot$, when the single-mode template works well for detection
purposes, the bias on frequency and quality factor has opposite sign depending
on whether the signals are in phase ($\phi_1=\phi_2=0$) or dephased
($\phi_1=0$, $\phi_2=\pi$). This is no coincidence, as we will demonstrate
below by studying the dependence of the bias on the phase angles
$(\phi_1\,,\phi_2)$.

\subsection*{Effect of the relative phase of the modes on detection and parameter estimation}

So far we computed the FF assuming $\phi_1=0$ (so that the dominant QNM has
maximum amplitude at $t=0$). We only explored the effect of the relative QNM
phase by changing the sign\footnote{For $\phi_1=0$, a sign change in ${\cal
    A}$ is obviously equivalent to setting $\phi_2=2n\pi$ (what we referred to
  as the QNM signals being ``in phase'') or $\phi_2=(2n+1)\pi$ (``dephased''
  signals), with $n$ an integer.} of ${\cal A}$. In practice the situation is
not so simple, since the phase of the two (or more) components of the
``exact'' signal is not known in advance.  This problem is reminiscent of the
analogous problem occurring in matched-filtering detection of inspiral signals
(see Appendix B of Ref.~\cite{Damour:1997ub}). For both inspiral and ringdown,
maximizing the FF (\ref{ff}) over all parameters yields a ``best possible
overlap'' which is somewhat optimistic, and therefore not too useful as a
detectability criterion. More realistically, we should take into account our
ignorance of the phase, or phases, of the ``true'' signal. This can be done by
computing a ``minimax'' overlap \cite{Damour:1997ub}: first maximize the FF
(\ref{ff}) over all parameters $\{\vec \lambda\}$ of the template $T\{\vec
\lambda\}$, and then minimize over the unknown phases of the actual inspiral
or ringdown signal $h(t)$.

To discuss the difference between ``best'' and ``minimax'' overlaps in terms
of detection and parameter estimation, here we perform FF calculations in the
$(\phi_1\,,\phi_2)$ plane. We consider, for illustration, two cases:

\begin{itemize}
\item[(i)] An IMBH with $M_0=100~M_\odot$ as observed by LIGO;

\item[(ii)] An IMBH with $M_0=200~M_\odot$ as observed by Advanced LIGO.
\end{itemize}

The FF for $\phi_1=\phi_2=0$ in these two cases is marked by a circle and a
square, respectively, in Fig.~\ref{fig:ffLIGO}.  As in the rest of this
Section, our signal will be given by the two-mode waveform (\ref{twomode})
with ${\cal A}=0.3$ and $j=0.6$.  This simple model is sufficient for our
present purpose. A more detailed analysis (taking into account details of the
angular dependence of the radiation, including a better model of the ringdown
signal for spinning mergers, and possibly using waveforms from numerical
relativity) is a topic for future work.

\begin{figure}[ht]
\epsfig{file=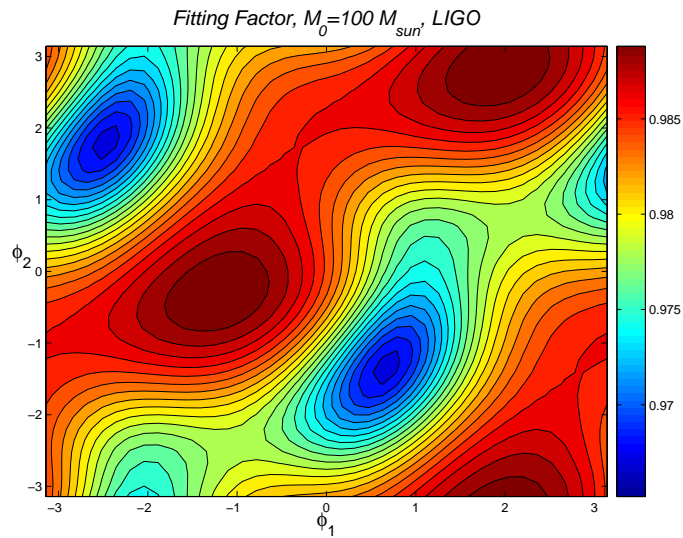,width=9cm,angle=0}
\caption{FF as a function of the phase angles of the signal in case (i).}
\label{fig:FFcontours}
\end{figure}

Fig.~\ref{fig:FFcontours} shows a contour plot of the FF for case (i). The FF
is always larger than $0.965$, or equivalently, the event loss is always
smaller than $10\%$. For this particular system (and, presumably, for most
low-mass black hole ringdowns) our ignorance of the phases does not sensibly
reduce our chances of detecting the signal.

The plot shows interesting features, some of which are easy to understand.
Under a simultaneous replacement $(\phi_1\,,\phi_2)\to
(\phi_1+(2n+1)\pi\,,\phi_2+(2m+1)\pi)$, where $(n\,,m)$ are integers, the FF
is unchanged. This is a trivial consequence of the fact that the overall sign
of the signal (\ref{twomode}) does not affect calculations of the FF.  The FF
has maxima and minima as a function of the two phases. Modulo periodicity, we
see that the minimum occurs when $\phi_1\sim \pi/4$ and $\phi_2\sim -\pi/4$.
This is reasonable: for these values of the phases the signals have comparable
magnitude and opposite phase at $t=0$, so that the $l=m=3$ multipole almost
exactly cancels out the dominant $l=m=2$ multipole in the initial (and
stronger) part of the signal.  This destructive interference produces a signal
that is sensibly different from a damped sinusoid, reducing the performance of
a simple single-mode filter.

\begin{figure}[ht]
\begin{tabular}{ll}
\epsfig{file=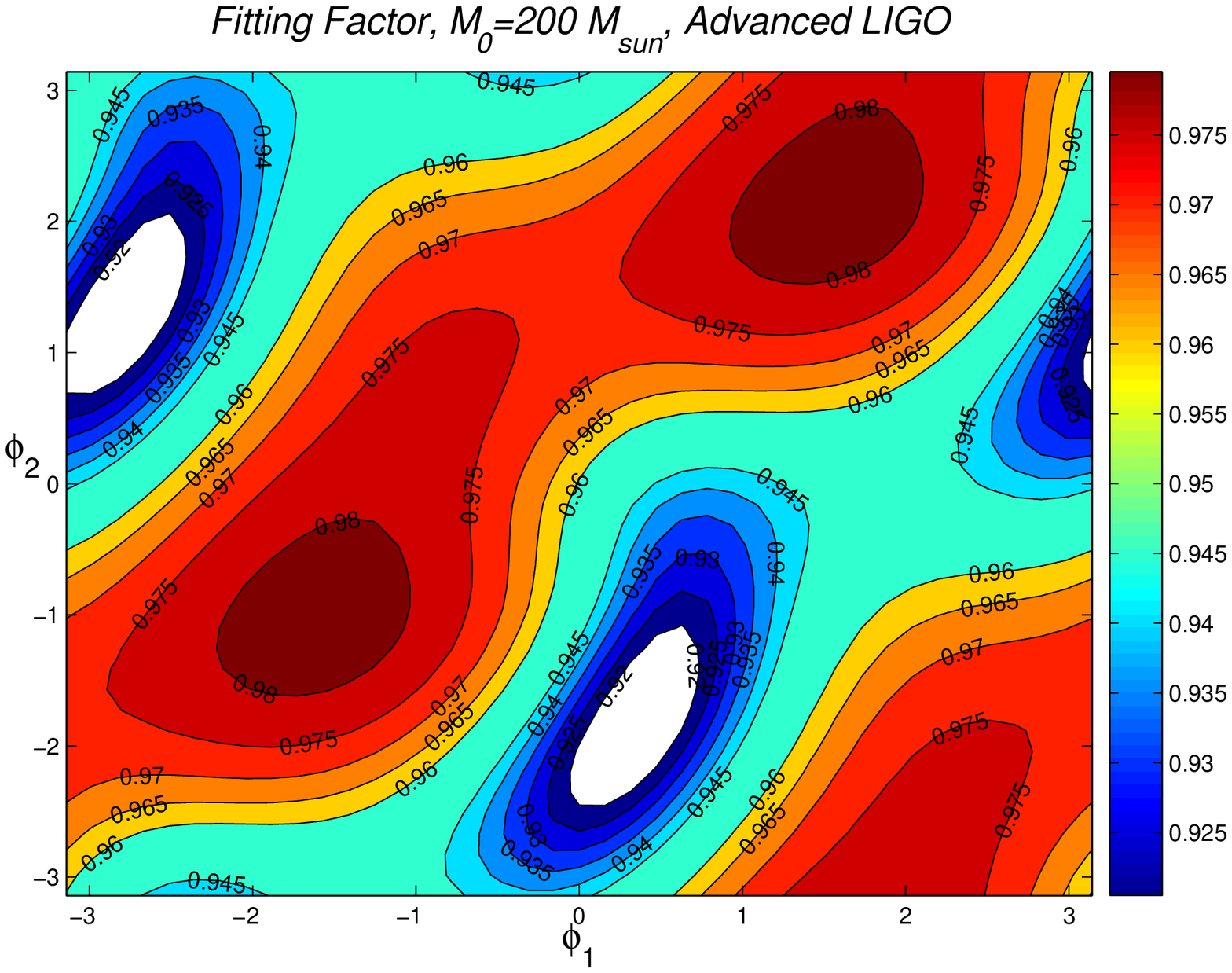,width=9cm,angle=0}&
\epsfig{file=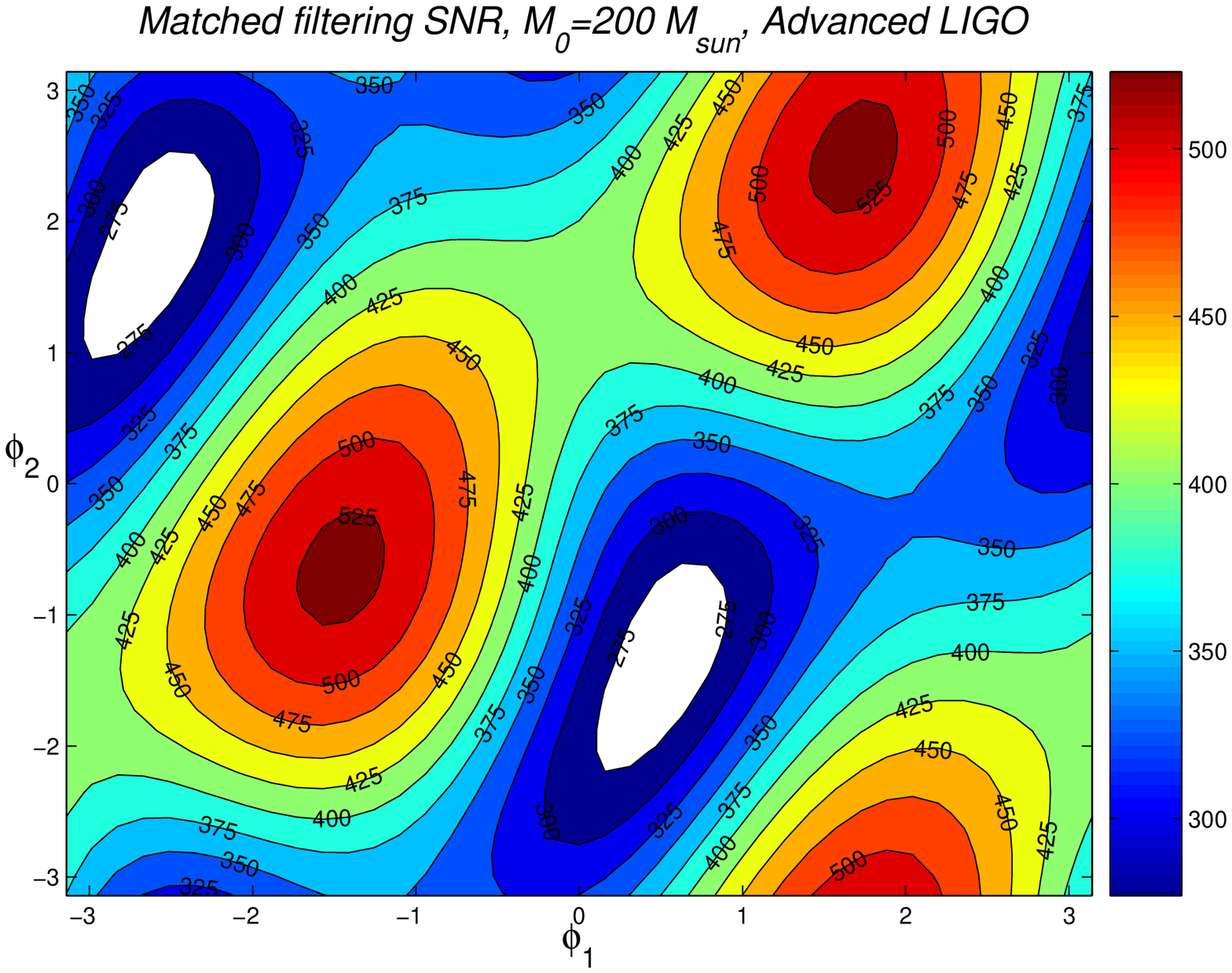,width=9cm,angle=0}
\end{tabular}
\caption{FF (left) and matched-filtering SNR (right) as a function of the
  phase angles of the signal in case (ii).}
\label{fig:ELcontours}
\end{figure}

Fig.~\ref{fig:ELcontours} shows contour plots of the FF and of the
matched-filtering SNR $\rho_{\rm MF}$, computed according to
Eq.~(\ref{rhomf}), for case (ii).  Now the FF is smaller than $0.965$ in
roughly half of the $(\phi_1\,,\phi_2)$ plane.  The event loss ranges from
$\sim 6\%$ to $\sim 22\%$, being larger than $10\%$ in roughly half of the
parameter space. From Fig.~\ref{fig:ffLIGO} we can expect that the event loss
would have been even larger if we had chosen larger values of the black hole
mass. The matched-filtering SNR was computed assuming that the overall
amplitude of the signal ${\cal A}_1$ corresponds to a ringdown efficiency
$\epsilon_{\rm rd}=3\%$, and that the luminosity distance $D_L=100$~Mpc. Both
the FF and the SNR show the (by now familiar) $\pi$-periodic pattern as a
function of the phase angles.

\begin{figure}[ht]
\begin{center}
\begin{tabular}{ll}
\epsfig{file=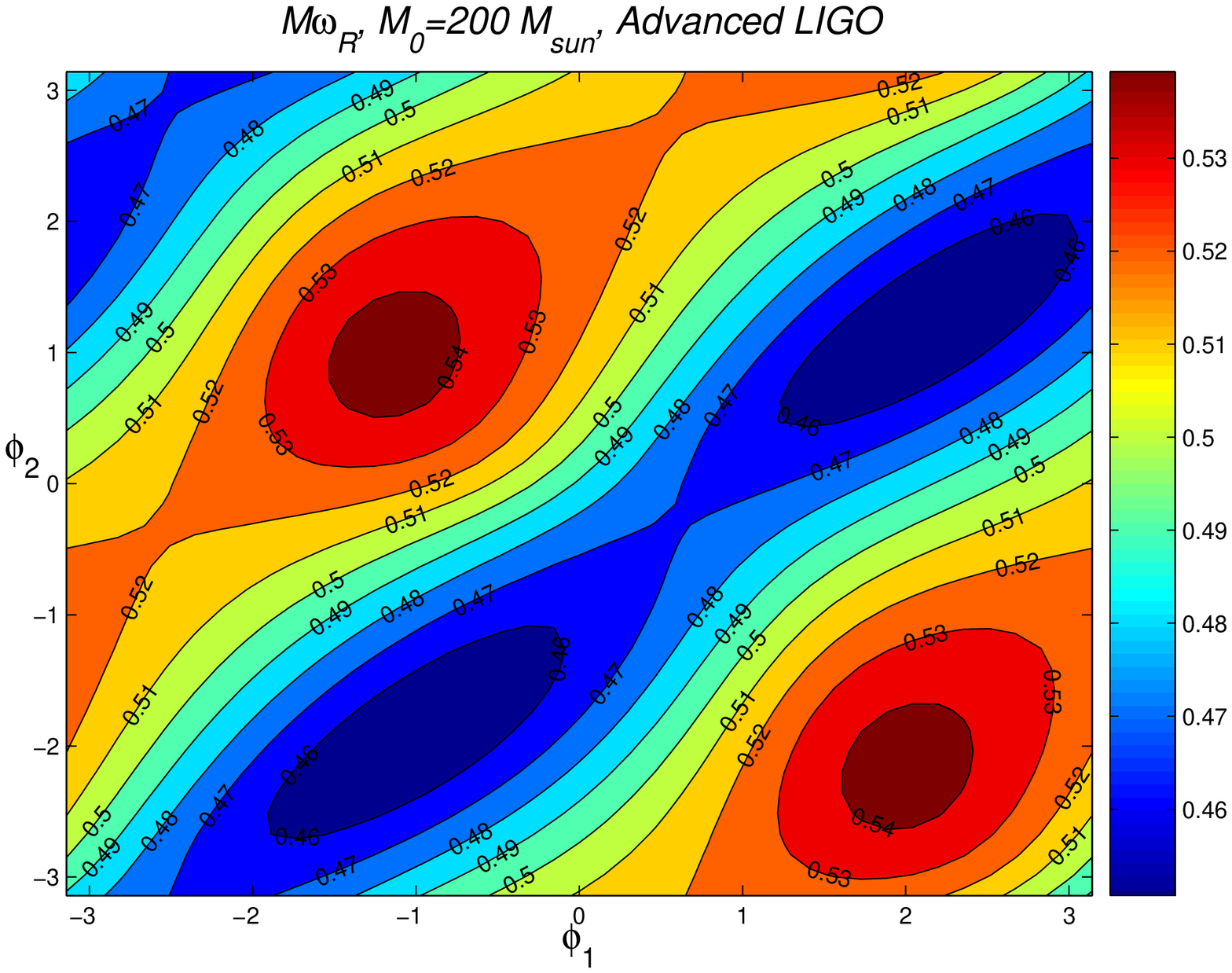,width=9cm,angle=0}&
\epsfig{file=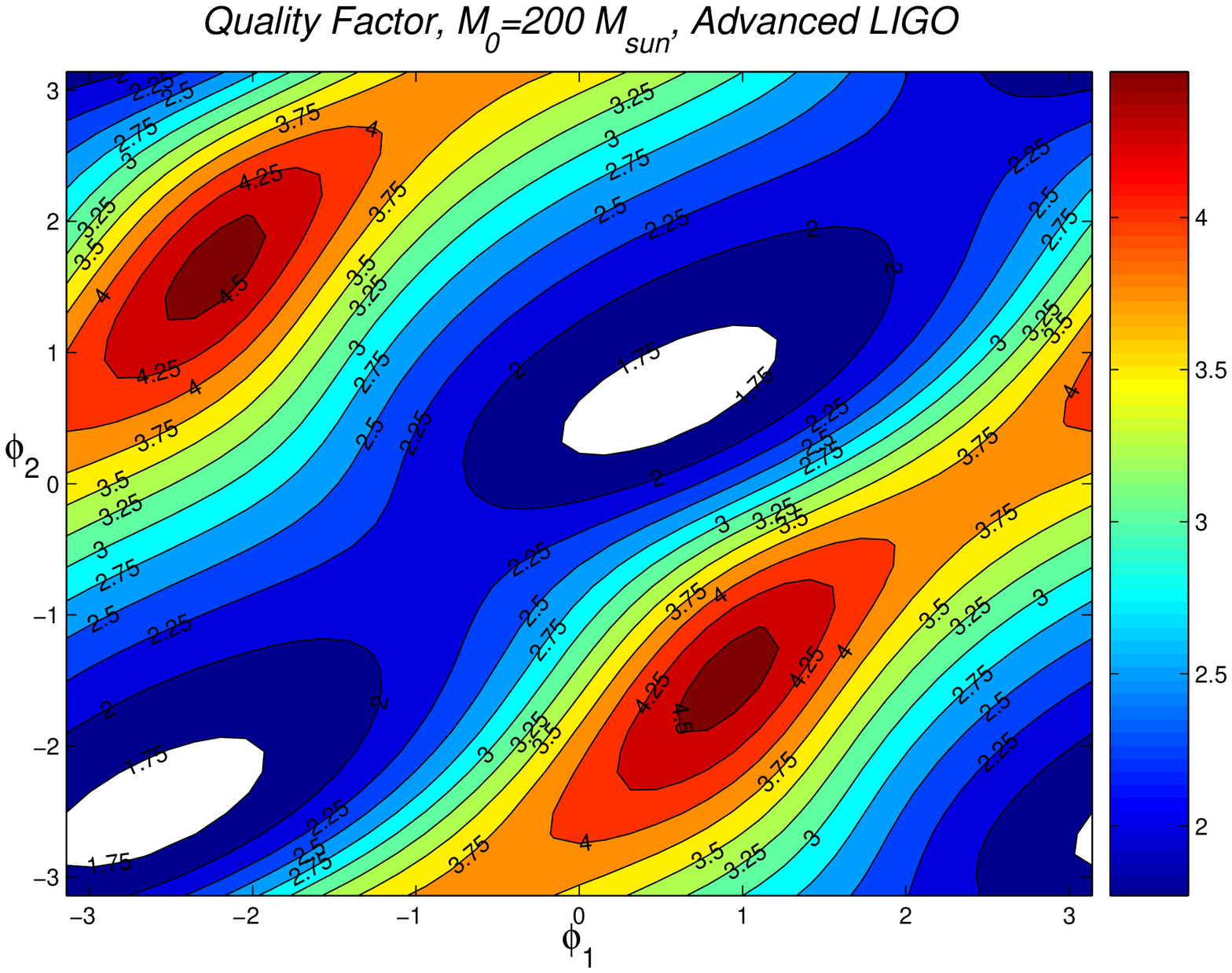,width=9cm,angle=0}\\
\end{tabular}
\end{center}
\caption{Dimensionless frequency (left) and quality factor (right) estimated
  by a single-mode filter in case (ii). The ``true'' frequencies and quality
  factors for a Kerr parameter $j=0.6$ are: $M\omega_{R\,1}=0.4940$,
  $Q_1=2.9490$, $M\omega_{R\,2}=0.7862$, $Q_2=4.5507$.
}
\label{fig:SNRQcontours}
\end{figure}

Suppose that the event loss is moderately large but not so large to prevent a
detection, as in case (ii). Then we may ask the question: what is the bias in
measured parameters when $(\phi_1\,,\phi_2)$ maximize the probability of
detection? In other words: when the template's frequency and quality factor
maximize the FF, do they also correspond to the ``true'' frequency and quality
factor of the $l=m=2$ fundamental mode?  Unfortunately, the answer is no.

Fig.~\ref{fig:SNRQcontours} shows the estimated dimensionless frequency (left)
and quality factor (right) as functions of the phase angles. The estimated
frequency has relatively small bias, and it always corresponds to the
least-damped mode in the pair. Results are more interesting for the quality
factor. For our chosen value of the Kerr parameter ($j=0.6$) the quality
factor of the $l=m=2$ and $l=m=3$ modes are $Q_1=2.9490$ and $Q_2=4.5507$,
respectively.  Comparing with Fig.~\ref{fig:FFcontours}, we see that relative
minima in $\rho_{\rm MF}$ (and in the FF) occur when the quality factor ``best
fits'' the $l=m=3$ mode.  This is a rather remarkable result: the filter
corresponding to the minimax overlap has a quality factor that ``best fits''
the subdominant mode in the pair. Unfortunately, maxima in $\rho_{\rm MF}$ do
{\it not} correspond to the filter being optimally adapted to the $l=m=2$
mode.  As the filter tries to maximize the overlap (and the SNR) the estimated
value of the quality factor becomes significantly biased, and it deviates
quite sensibly from the value expected from the dominant ($l=m=2$) mode. The
bottom line is, once again, that single-mode filters can be useful for
detection, but a multi-mode post-processing will be necessary for accurate
measurements of the black hole's angular momentum.

Another argument in favor of a two-stage search strategy comes from estimating
the number of multi-mode templates that would be necessary for a detection.
Searching for a larger number of modes implies a larger number of free
parameters, and a correspondingly larger bank of filters. Matched-filtering
requires that one covers the possible parameter space with a sufficiently fine
template mesh, so that the best-matching template lies close enough to the
true waveform. The distance between templates can be quantified in terms of a
metric in template space, that was introduced by Owen \cite{Owen:1995tm}
following Refs.~\cite{Sathyaprakash:1991mt,Apostolatos:1995pj}.  If the mesh
is too fine a very large number of templates may be required, a
computationally expensive option.  On the other hand, if the mesh is very
coarse the template lying ``closer'' to the true waveform may produce a large
event loss.  A multi-mode search increases the number of templates needed for
the mesh to be sufficiently fine.  For a single-mode search, Creighton
\cite{Creighton:1999pm} estimated that $\sim 500$ templates are necessary to
reduce the event loss below $\sim 10\%$ for LIGO and VIRGO. The same estimate
can be used to show that $\sim 1000$ templates would be necessary for
single-mode templates with LISA \cite{Berti:2006ew}. In Appendix
\ref{app:numbertemplates} we estimate that the number of filters ${\cal N}$
required for a two-mode template search would be much larger:
\beq
{\cal N}
&\approx& b \times 10^6 \left (\frac{0.03}{1-{\rm MM}}\right )^{5/2}\,, \eeq
where $b$ is a factor of order unity which depends on the detector's frequency
span: with our choice of $f_s$ we get $b=8.3\,,2.2\,,1.6\,,1.2\,,1.6$ for
LISA, EGO, Advanced LIGO, LIGO and Virgo, respectively. Using better template
placing techniques \cite{nakano} or imposing constraints on the functional
form of the QNM frequencies (Appendix \ref{app:numbertemplates}) could help
reduce computational requirements. A two-stage search seems to be a good
compromise between performance and computational costs.  Single-mode templates
can be used for detection. Given a detection, multi-mode templates or Prony
methods \cite{bcgs} should be used for parameter estimation. A larger number
of templates also means that the threshold for detection must be set higher,
because there is a larger false alarm probability. Hierarchical searches or
other techniques could play an important role in this regard
\cite{reductionfalsealarm}. In any case, such a large number of templates may
not be a problem by the time advanced detectors are in operation.  Improved
computer performance and the use of large-scale computational projects, such
as Einstein@Home \cite{einsteinathome}, could be sufficient to overcome
computational difficulties within the next decade.

\section{\label{sec:resolvability} Mode resolvability and no-hair tests}

So far we looked at the event loss and bias in parameter estimation due to the
use of single-mode templates to detect multi-mode signals. The discussion
assumed that the gravitational wave signal is composed of at least two QNMs
having roughly comparable amplitude. The question we address here is the
following: how can we tell if there really are two or more modes in the
signal, and can we resolve their parameters? If the noise is large and the
amplitude of the weaker signal is very low, or the two signals have almost
identical frequencies, the two modes could be difficult to resolve. This issue
is particularly significant since no-hair tests using ringdown \cite{bcw}
require the presence and resolvability of two or more modes. Roughly speaking,
the first mode is used to measure $M$ and $j$ by inverting the relations
$f_1=f_1(M\,,j)\,,Q_1=Q_1(M\,,j)$ (see Appendix \ref{app:numbertemplates} for
more details); the second mode can then be used to test consistency with the
Kerr geometry.

Here we really address two different issues. We first assume that there are
indeed two modes in the signal, and we discuss criteria to resolve their {\it
  frequencies and damping times}. This discussion parallels that in
Ref.~\cite{bcw}, updating estimates of the relative mode excitation on the
basis of recent results from numerical relativity, and correcting a typo in
that paper.  Then we introduce a rigorous criterion to resolve {\it
  amplitudes}: that is, we compute the minimum SNR such that one can decide by
the presence of two modes in a given ringdown signal.

{\it Resolving frequencies and damping times.}
A crude lower limit on the SNR required to resolve frequencies and damping
times was presented in Ref.~\cite{bcw}. The analysis uses the statistical
uncertainty in the determination of each frequency and damping time, which a
standard Fisher Matrix calculation\footnote{Our Eq.~(\ref{sigmaffh}) corrects
  a missing factor of $2\pi$ in Ref.~\cite{bcw}.} estimates to be \cite{bcw}
\begin{subequations}
\beq
\rho \sigma_{f_1}&=&
\f{\pi}{\sqrt{2}}
\left\{
\f{f_1^3\left(3+16Q_1^4\right)}{{\cal A}_1^2 Q_1^7}
\left[
\f{{\cal A}_1^2 Q_1^3}{f_1\left(1+4Q_1^2\right)}+
\f{{\cal A}_2^2 Q_2^3}{f_2\left(1+4Q_2^2\right)}
\right] \right\}^{1/2}\,,\label{sigmaffh}\\
\rho \sigma_{\tau_1}&=& \f{2}{\pi} \left\{ \f{\left(3+4Q_1^2\right)}{{\cal A}_1^2 f_1 Q_1} \left[ \f{{\cal
A}_1^2 Q_1^3}{f_1\left(1+4Q_1^2\right)}+ \f{{\cal A}_2^2 Q_2^3}{f_2\left(1+4Q_2^2\right)} \right]
\right\}^{1/2}\,. \eeq
\end{subequations}
These errors refer to mode ``1'' in a pair. By considering the ``symmetric''
case $\phi_1=\phi_2=0$, the errors on $f_2$ and $\tau_2$ are simply obtained
by exchanging indices ($1\leftrightarrow 2$). The expression above holds in
both the FH and EF conventions, assuming white-noise for the detector, but
modes ``1'' and ``2'' must correspond to different values of $l$ or $m$ (in
the nomenclature used in Ref.~\cite{bcw}, the QNMs must be quasi-orthonormal).

A natural criterion ({\it \'a la} Rayleigh) to resolve frequencies and damping
times is
\be\label{criterion} |f_1-f_2|>{\rm max}(\sigma_{f_1},\sigma_{f_2})\,,\qquad |\tau_1-\tau_2|>{\rm
max}(\sigma_{\tau_1},\sigma_{\tau_2})\,. \ee
In interferometry this would mean that two objects are (barely) resolvable if
``the maximum of the diffraction pattern of object 1 is located at the minimum
of the diffraction pattern of object 2''. We can introduce two ``critical''
SNRs required to resolve frequencies and damping times,
\be
\rho_{\rm crit}^f =
\f{{\rm max}(\rho \sigma_{f_1},\rho \sigma_{f_2})}{|f_1-f_2|}\,,\qquad
\rho_{\rm crit}^\tau = \f{{\rm max}(\rho \sigma_{\tau_1},\rho
\sigma_{\tau_2})}{|\tau_1-\tau_2|}\,,
\ee
and recast our resolvability conditions as
\begin{subequations}
\label{minimal}
\beq
\rho&>&
\rho_{\rm crit}
={\rm min}(\rho_{\rm crit}^f,\rho_{\rm crit}^\tau)\,,\\
\label{both}
\rho&>&
\rho_{\rm both}
={\rm max}(\rho_{\rm crit}^f,\rho_{\rm crit}^\tau)\,. \eeq
\end{subequations}
The first condition implies resolvability of either the frequency or the
damping time, the second implies resolvability of both.

{\it Resolving amplitudes.}
A related question is: how large a SNR do we need to confidently say that we
have detected a multi-mode signal, and to resolve two signals of different
amplitudes? Suppose again, for simplicity, that the true signal is a two-mode
superposition. Then we expect the weaker signal to be hard to resolve if its
amplitude is low and the detector's noise is large.

In Appendix \ref{app:amplres} we quantify this statement by deriving a
critical SNR for amplitude resolvability based on the generalized likelihood
ratio test (GLRT), $\rho_{\rm GLRT}$. The derivation of this critical SNR,
which is given in Eq.~(\ref{eq:cc}), is based on the following simplifying
assumptions: (i) using other criteria, we have already decided for the
presence of one ringdown signal, and (ii) the parameters of the ringdown
signal (frequencies and damping times), as well as the amplitude of the
dominant mode, are known. In practice the latter assumption is not valid. For
this reason, our estimates of the minimum SNR needed to detect more than one
mode should be considered optimistic.

\begin{figure*}[ht]
\epsfig{file=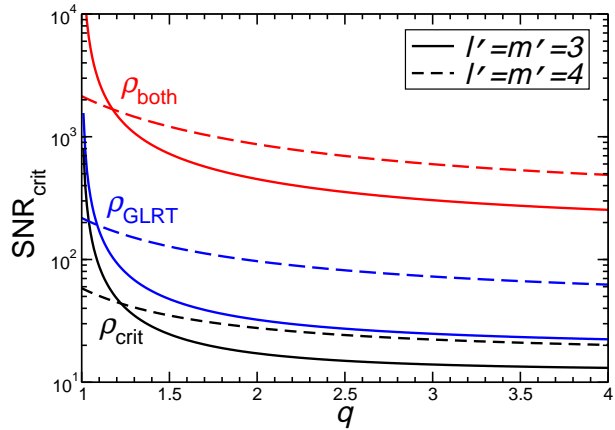,width=7cm,angle=-90} 
\caption{Minimum SNR required to resolve two modes, as function of the
  binary's mass ratio $q$. If $\rho>\rho_{\rm GLRT}$ we can tell the presence
  of a second mode in the waveform, if $\rho>\rho_{\rm crit}$ we can resolve
  either the frequency or the damping time, and if $\rho>\rho_{\rm both}$ we
  can resolve both. Mode ``1'' is assumed to be the fundamental mode with
  $l=m=2$; mode ``2'' is either the fundamental mode with $l=m=3$ (solid
  lines) or the fundamental mode with $l=m=4$ (dashed lines).}
\label{fig:minimumSNR}
\end{figure*}

Fig.~\ref{fig:minimumSNR} compares the critical SNR $\rho_{\rm GLRT}$, as
defined in Eq.~(\ref{eq:cc}), and the two different criteria for frequency
resolvability of Eq.~(\ref{minimal}).  All quantities are computed as
functions of the binary's mass ratio $q$. The angular momentum $j$ of the
final black hole is computed using the fitting formula derived in
Ref.~\cite{Berti:2007fi} for the angular momentum of the black hole resulting
from unequal-mass, nonspinning binary black hole mergers:
$j=3.352\eta-2.461\eta^2$, where the symmetric mass ratio $\eta\equiv
q/(1+q)^2$. This value of $j$ is then used to read QNM frequencies from
numerical tables. We assume the dominant mode to be the fundamental $l=m=2$
QNM and for the subdominant mode we take the fundamental QNMs with $l=m=3$ or
$l=m=4$.  To compute the relative amplitude ${\cal A}(q)$ for different mass
ratios we use the EMOP estimate of Eq.~(\ref{Aq}).

The plot shows that $\rho_{\rm crit}<\rho_{\rm GLRT}<\rho_{\rm both}$ for all
values of $q$. Therefore, given a detection, the most important criterion to
determine whether we can carry out no-hair tests seems to be the GLRT
criterion. If $\rho>\rho_{\rm GLRT}$ we can decide for the presence of a
second mode in the signal. Whenever the second mode is present, we also have
$\rho>\rho_{\rm crit}$: that is, we can resolve at least the frequencies (if
not also the damping times) of the two modes. A SNR $\rho \sim 30-40$ is
typically enough to perform the GLRT test on the $l=m=3$ mode, as long as
$q\gtrsim 1.5$ or so (equal-mass mergers should be quite rare anyway). By
looking at Fig.~\ref{fig:snr} we conclude that not only LISA, but also
advanced Earth-based detectors (Advanced LIGO and EGO) have the potential to
identify Kerr black holes as the vacuum solutions of Einstein's general
relativity.

\section{Conclusions}
\label{conclusions}

In this paper we analyze the detectability of ringdown waves by Earth-based
interferometers. Confirming and extending previous analyses, we show that
Advanced LIGO and EGO could detect intermediate-mass black holes of mass up to
$\sim 10^3~M_\odot$ out to a luminosity distance of a few Gpc.

Using recent results for the multipolar energy distribution from numerical
relativity simulations of non-spinning binary black hole mergers
\cite{Berti:2007fi} to estimate the relative amplitude of the dominant
multipolar components, we point out that the single-mode templates presently
used for ringdown searches in the LIGO data stream could produce a significant
event loss ($> 10\%$ in a large interval of black hole masses). A similar
event loss should affect also next-generation Earth-based detectors, as well
as the planned space-based interferometer LISA.

Single-mode templates are useful for detection of low-mass systems, but they
produce large errors in the estimated values of the parameters (and especially
of the quality factor). We estimate that, unfortunately, more than $\sim 10^6$
templates would be needed for a single-stage multi-mode search.  For this
reason we recommend a ``two stage'' search to save on computational costs: a
single-mode template could be used to detect the signal, and a multi-mode
template (or even better, Prony methods \cite{bcgs}) could be used to estimate
parameters once a detection has been made.

In Appendix \ref{app:amplres} we introduce a criterion to decide for the
presence of more than one mode in a ringdown signal. By updating estimates of
the critical signal-to-noise ratio required to resolve the frequencies of
different QNMs using results from numerical relativity, we show that
second-generation Earth-based detectors and LISA both have the potential to
perform tests of the Kerr nature of astrophysical black holes.

In the future we plan to use numerical waveforms (possibly including spin
effects) to refine our estimates. We also plan to carry out Monte Carlo
simulations to study the information that can be extracted on the source
position and orientation using a network of Earth-based detectors. The
possibility to constrain the black hole spin's direction from the multipolar
distribution of the merger-ringdown radiation should be particularly
interesting (eg. for coincident electromagnetic observations of jets that
could be emitted along the black hole spin's axis).

\section*{Acknowledgements}

We are grateful to Alessandra Buonanno, Kostas Kokkotas, Clifford Will and
Nicolas Yunes for discussions. This work was partially funded by Funda\c c\~ao
para a Ci\^encia e Tecnologia (FCT) –- Portugal through projects
PTDC/FIS/64175/2006 and POCI/FP/81915/2007, by the National Science Foundation
under grant numbers PHY 03-53180 and PHY 06-52448, and by NASA under grant
number NNG06GI60 to Washington University.

\clearpage

\appendix

\section{Number of templates for a two-mode search}
\label{app:numbertemplates}

Two-mode ringdown waveforms of the form (\ref{twomode}) depend of five
parameters: two quality factors $Q_1\,,Q_2$, two frequencies $f_1\,,f_2$ and
one relative amplitude ${\cal A}$ between the different modes (for simplicity,
here we set $\phi_1=\phi_2=0$).
For matched-filtering with ringdown waveforms of unknown frequency and quality
factor we must lay down a ``mesh'' covering the parameter space with some
pre-defined precision (that can be translated to a pre-defined minimum loss of
SNR)
\cite{Sathyaprakash:1991mt,Owen:1995tm,Creighton:1999pm}.
We follow Owen's formalism \cite{Owen:1995tm} to estimate the necessary number
of templates. The distance between nearby templates, which defines the
mismatch between the filters, can be computed in terms of the metric
\be ds^2 =g_{\mu \nu}dx^{\mu}dx^{\nu}\,,\quad
x^{\mu}=(Q_1\,,Q_2\,,f_1\,,f_2\,,{\cal A})\,, \ee
where (for large or moderate values of $Q_1$ and $Q_2$) the metric
coefficients are well approximated by
\begin{subequations}
\begin{widetext}
\beq g_{Q_1Q_1}&=&\frac{f_2\left (f_2Q_1+2{\cal A}^2f_1Q_2\right )} {8Q_1\left
(f_2Q_1+{\cal A}^2f_1Q_2\right )^2}\,,\quad g_{Q_2Q_2}=\frac{{\cal A}^2f_1\left
(2f_2Q_1+{\cal A}^2f_1Q_2\right )}
{8Q_2\left (f_2Q_1+{\cal A}^2f_1Q_2\right )^2}\,,\\
g_{f_1f_1}&=&\frac{f_2Q_1^3}{f^{(1)\,2}\left (f_2Q_1+{\cal A}^2f_1Q_2\right
)}\,,\quad g_{f_2f_2}=\frac{{\cal A}^2f_1Q_2^3}{f^{(2)\,2}
\left (f_2Q_1+{\cal A}^2f_1Q_2\right )}\,,\\
g_{{\cal A}{\cal A}}&=&\frac{f_1f_2Q_1Q_2}{2\left (f_2Q_1+{\cal A}^2f_1Q_2\right )^2}\,,\quad
g_{Q_1Q_2}=-\frac{{\cal A}^2f_1f_2}{8\left (f_2Q_1+{\cal A}^2f_1Q_2\right )^2}\,,\\
g_{Q_1f_1}&=&-\frac{f_2\left (f_2Q_1+2{\cal A}^2f_1Q_2\right )} {8f_1\left
(f_2Q_1+{\cal A}^2f_1Q_2\right )^2}\,,\quad
g_{Q_1f_2}=\frac{{\cal A}^2f_1Q_2}{8\left (f_2Q_1+{\cal A}^2f_1Q_2\right )^2}\,,\\
g_{Q_2f_1}&=&\frac{{\cal A}^2f_2Q_1}{8\left (f_2Q_1+{\cal A}^2f_1Q_2\right )^2}\,,\quad
g_{Q_2f_2}=-\frac{{\cal A}^2f_1\left (2f_2Q_1+{\cal A}^2f_1Q_2\right )}
{8f_2\left (f_2Q_1+{\cal A}^2f_1Q_2\right )^2}\,,\\
g_{Q_1{\cal A}}&=&-\frac{{\cal A}f_1f_2Q_2}{4\left (f_2Q_1+{\cal A}^2f_1Q_2\right )^2}\,,\quad
g_{Q_2{\cal A}}=\frac{{\cal A}f_1f_2Q_1}{4\left (f_2Q_1+{\cal A}^2f_1Q_2\right )^2}\,,\\
g_{f_1f_2}&=&-\frac{{\cal A}^2Q_1Q_2}{8\left (f_2Q_1+{\cal A}^2f_1Q_2\right )^2}\,,\quad
g_{f_1{\cal A}}=\frac{{\cal A}f_2Q_1Q_2}{4\left (f_2Q_1+{\cal A}^2f_1Q_2\right )^2}\,,\\
g_{f_2{\cal A}}&=&-\frac{{\cal A}f_1Q_1Q_2}{4\left (f_2Q_1+{\cal A}^2f_1Q_2\right )^2} \,.
\eeq
\end{widetext}
\end{subequations}
Requiring a loss of no more than $10\%$ in the event rate due to a mismatched
template (i.e, the {\it minimal match} MM \cite{Owen:1995tm,Creighton:1999pm}
should be at least $0.97$) we get an estimate for the number of templates we
need:
\be
{\cal N}=\frac{\int dQ_1dQ_2df_1df_2d{\cal A} \sqrt{{\rm det}||g_{\mu\nu}||}}
{32\left[(1-{\rm MM})/5\right]^{5/2}}
\approx b \times 10^6 \left (\frac{0.03}{1-{\rm MM}}\right )^{5/2}\,.
\ee
Here $b$ is a factor of order unity which depends on the detector's frequency
span. We get $b=8.3\,,2.2\,,1.6\,,1.2\,,1.6$ for LISA, EGO, Advanced LIGO,
LIGO and Virgo, respectively.  In deriving this number we assume the
frequencies to be searched for are those of interest for each of the detectors
($3\times 10^{-5}\lesssim f \lesssim 1$ for LISA, $10 \lesssim f \lesssim
2000$ for EGO, $20 \lesssim f \lesssim 2000$ for Advanced LIGO and Virgo,
$40\lesssim f \lesssim 2000$ for LIGO), that the quality factor varies between
$0$ and $20$ for all modes likely to be detected \cite{bcw}, and that the
relative amplitude ${\cal A}$ varies between $0.01$ and $100$. Our estimates
are not strongly sensitive to the relative amplitude: assuming $0<{\cal A}<1$
yields a total number of templates which is roughly half the above number.

For the single-mode case, setting $Q_2=f_2={\cal A}=0$ we get the following
metric:
\be ds^2 \approx \frac{1}{8 Q^2}dQ^2-\frac{1}{4Qf}dQdf+\frac{Q^2}{f^2}df^2\,. \ee
The number of templates ${\cal N} \sim 6 \,Q_{\rm max}\log{(f_{\rm max}/f_{\rm
    min})}$ \cite{Creighton:1999pm}. In particular we get ${\cal N} \sim 460$
for LIGO and Virgo \cite{Creighton:1999pm}, and ${\cal N} \sim 1000$ for LISA
\cite{Berti:2006ew} (a huge improvement in terms of computational
requirements).

The formalism and numbers presented in this section are valid for a general
ringdown signal: no constraints were imposed on the QNM spectrum. A possible
approach to reduce the number of templates is to {\it assume} that the source
is a general relativistic black hole. In this case, we are left with only
three intrinsic parameters: the mass and angular momentum of the black hole
and the relative amplitude between the modes. Alternatively, we can choose the
independent parameters to be one quality factor $Q_1$, one frequency $f_1$ and
the relative amplitude of the modes; the quality factor and frequency of the
second mode, $Q_2$ and $f_2$, can be thought of as functions of $Q_1$ and
$f_1$.

More explicitly, we find that for rotations $0\leqslant j\leqslant0.98$ the
frequencies and quality factors of the fundamental mode with $l=m=2,~3,~4$ are
well approximated (to within $\sim 6\%$ or better) by
\beq 2\pi M f_{ll0}&\approx & 0.74+0.39l-\left (0.78+0.18l \right )\left (1-j\right )^{\frac{3.2+4.9l}{100}}\,,\\
Q_{ll0}&\approx& 0.26+0.22l+\left (-0.36+0.88l \right )\left (1-j\right )^{-0.49}\,.
\eeq
By inverting these relations for (say) the $l=m=2$ mode, one can infer
$(j,~M)$ and compute the frequency and quality factor of the modes with
$l=m=3,~4$. More generally, in Ref.~\cite{bcw} the frequencies and quality
factors of the first three overtones (for $l=2,~3,~4$ and all values of $m$)
were fitted by functions of the form
\beq
2\pi M f_{lmn}&=&f_1+f_2(1-j)^{f_3}\,,\label{ffit}\\
Q_{lmn}&=&q_1+q_2(1-j)^{q_3}\,.\label{Qfit}\eeq
Here the constants $f_i$ and $q_i$ depend on $(l\,,m\,,n)$ (see Tables VIII-X
in \cite{bcw}), and the fits are accurate to better than $4\%$. By using these
fits or (more precisely) numerical QNM data we can express any subdominant
mode in terms of the dominant mode, and possibly reduce the number of
templates.

Another possibility to reduce the number of templates is to restrict the
parameter space by using information derived from measurements of the {\it
  inspiral} waveforms. For example, if we had a reasonably accurate
measurement of the masses and spins of the binary members we could
significantly restrict the possible values of the mass and spin of the final
black hole to be searched for.

\section{Amplitude resolvability}
\label{app:amplres}

The purpose of this Appendix is to estimate the minimum SNR required to test
the hypothesis that a second mode is present in a ringdown waveform. The
derivation is based on the following simplifying assumptions: (i) using other
criteria, we have already decided for the presence of one (dominant) damped
sinusoid in the signal, and (ii) the parameters of the ringdown signal
(frequencies and damping times), as well as the amplitude of the dominant
mode, are known. In practice the latter assumption is not valid, so our
estimates of the minimum SNR should be considered optimistic.

The question of whether one or two damped sinusoids are present in the signal
can be stated in statistical terms, as follows\footnote{See also the work by
  Milanfar and Shahram \cite{smilanfar,milanfars}.}. Let $w(t)$ be a zero-mean
gaussian white noise process, and define $y(t) \equiv s(t) - {\cal A}_1
h_1(t)$ to be the difference between the actual signal $s(t)$ and the dominant
normalized QNM signal $h_1(t)= e^{-(\pi f_1/Q_1) t} \sin(2\pi f_1 t)$ (for
simplicity, in the present discussion we set $\phi_1=\phi_2=0$).  Denote by
${\cal H}_1$ the hypothesis that the signal contains only one damped
exponential in noise, and by ${\cal H}_2$ the hypothesis that the signal
consists of two damped exponentials in noise:
\be
\begin{cases}
{\cal H}_1: y(t) = w(t)\\
{\cal H}_2: y(t) = {\cal A}_2\,h_2(t) + w(t)
\end{cases}
\ee
We write the normalized second QNM as $h_2(t)=e^{-(\pi f_2/Q_2) t} \sin(2\pi
f_2 t)$.

We assume that we do not possess any prior information on the possible values
of ${\cal A}_2$, besides the fact that ${\cal A}_2>0$. The general structure
of composite hypothesis testing is involved when unknown parameters (in this
case, ${\cal A}_2$) appear in the probability density function.  We follow
Milanfar and Shahram \cite{smilanfar,milanfars} and consider the generalized
likelihood ratio test, which proceeds by computing first the maximum
likelihood (ML) estimates of the unknown parameters. These estimates are then
used to form Neyman-Pearson detectors. We note that this approach has been
compared very favorably against other standard tests in Ref.~\cite{smilanfar}.

Suppose that the signal $s(t)$ is sampled at discrete times
$t_k\,\,(k=1,\dots, N)$ and that the corresponding sample values are $s_k =
s(t_k),\, y_k \equiv s_k - {\cal A}_1 h_1(t_k)$. If we assume additive white
noise with variance $\sigma$, the probability of the observed data under the
hypothesis of a second damped sinusoid of amplitude ${\cal A}_2$ in the signal
is
\be p_{{\cal A}_2} = \prod_{i=1}^{N} p\left(y(x_i)\right) = \prod_{k=1}^{N} \frac{1}{\sigma\sqrt{2\pi}}
\exp{\left[-\f{(y_k-{\cal A}_2\,h_{2\,k})^2}{2\sigma^2}\right]} \,. \ee
To decide on hypothesis ${\cal H}_1$ or ${\cal H}_2$, we start by computing
the ML estimate of the unknown parameter ${\cal A}_2$:
\be \max_{{\cal A}_2} p_{{\cal A}_2} = \max_{{\cal A}_2} \ln p_{{\cal A}_2}= \min_{{\cal A}_2} \sum_{k = 1}^{N}
(y_k-{\cal A}_2\,h_{2\,k})^2 = \min_{{\cal A}_2} ||\textbf{Y}-{\cal A}_2\,\textbf{H}_2||^2\,,
\ee
where $\textbf{Y}$ and $\textbf{H}_2$ are column-vectors of the samples $y_k$
and $h_{2\,k}$, respectively, a superscript $T$ stands for ``transpose'', and
$||\textbf{V}||^2\equiv \textbf{V}^T \textbf{V}$ denotes the norm of a vector
$\textbf{V}$.  The (unconstrained) ML estimate $\hat{{\cal A}}_2$ of the
parameter ${\cal A}_2$ is then
\be \hat{{\cal A}}_2 = \f{\textbf{H}_2^T\textbf{Y}} {||\textbf{H}_2||^2}\,. \ee

We wish to test the hypothesis that ${\cal A}_2 > 0$ against the hypothesis
that ${\cal A}_2 = 0$.  There is no general-purpose test to do this, but a
powerful test is to compute a likelihood ratio and to find the maximum in both
the numerator and denominator: this is called the {\em generalized likelihood
  ratio test} (GLRT). Some algebra shows that
\be T(\textbf{Y})\equiv \ln \frac{\max_{{\cal H}_2} p_{{\cal A}_2}(y_x)}
{\max_{{\cal H}_1} p_{{\cal A}_2=0}(y_x)}
%
%
=
\f{1}{2}
\left(
\frac{\textbf{H}_2^T\textbf{Y}}{\sigma||\textbf{H}_2||}
\right)^2\,. \\
\ee
For any given data set $\textbf{Y}$, we decide on ${\cal H}_2$ if
$\sqrt{2T(\textbf{Y})}$ exceeds a specified threshold $\gamma$:
\be \f{\textbf{H}_2^T\textbf{Y}} {\sigma||\textbf{H}_2||} > \gamma \,.\ee
While it may seem troublesome to use the unconstrained ML estimate to form the
GLRT, in fact, due to the assumed positivity of ${\cal A}_2$, the detector
structure is effectively producing a one-sided test and therefore this is in
fact a uniformly most powerful detector \cite{Scharf1991}.

The choice of $\gamma$ is motivated by the level of tolerable false-positive
rate \cite{smilanfar,milanfars}.  The detection rate $P_d$ and false alarm
rate $P_f$ for this detector are related as
\be \label{eq:aa} P_d = Q ({\cal A}_2\,\eta+\gamma) = Q ({\cal A}_2\,\eta+Q^{-1}(P_f))\,, \ee
where
\begin{equation}
\label{eq:etaDefinition} \eta =
\f{||\textbf{H}_2||}{\sigma}
\,.
\end{equation}
Here $Q$ denotes the right-tail probability function for a standard Gaussian
random variable (zero mean and unit variance):
\be Q(x) = \int_{x}^\infty \frac{1}{\sqrt{2\pi}}
\exp{\left[-\frac{w^2}{2}\right]} dw\,. \ee
From \eqref{eq:aa} we find
\be \label{eq:BB} Q^{-1}(P_d) - Q^{-1}(P_f) =  {\cal A}_2\,\eta \,.\ee
This last expression can now be put in a more convenient form, by defining the
output SNR as
\be \rho = \frac{||{\cal A}_1\textbf{H}_1 + {\cal A}_2\textbf{H}_2||}{\sigma}\,. \ee
Using \eqref{eq:etaDefinition} and \eqref{eq:BB} the relation between the
minimum resolvable ${\cal A}={\cal A}_2/{\cal A}_1$ and the required SNR can
be made explicit. The critical SNR for hypothesis testing $\rho_{\rm GLRT}$
is
\be
\rho_{\rm GLRT} =
\left[Q^{-1}(P_d) - Q^{-1}(P_f)\right] \frac{||\textbf{H}_1 + {\cal
A}\textbf{H}_2||}{||{\cal A} \textbf{H}_2||} \,.
\label{eq:cc}
\ee
For the calculations in Fig.~\ref{fig:minimumSNR} we set $P_d=10^{-2}$,
$P_f=0.99$. Had we chosen $P_d=0.1$ and $P_f=0.9$, the critical $\rho$ would
have decreased by a factor $\sim 2.6/4.6$. For more stringent false alarm
rates (say, with $P_d=10^{-6}$ and $P_f=0.99$) the critical SNR would have
increased by a factor $\sim 7.1/4.6$.

\clearpage


\end{document}